\documentclass[a4j,11pt]{article}

\usepackage{amsmath,amssymb,color,graphics,graphicx,amscd,amsfonts,epsf}
\allowdisplaybreaks[4]

\date{}
\begin{document}

\begin{flushright} 
KUNS-2185\\
WIS/02/09-JAN-DPP\\
AEI-2009-003
\end{flushright} 

\vspace{0.1cm}

\begin{center}
  {\LARGE
On the shape of a D-brane bound state and its topology change
  }
\end{center}
\vspace{0.1cm}
\begin{center}

         Tatsuo A{\sc zeyanagi}$^{a}$\footnote
           {
E-mail address : aze@gauge.scphys.kyoto-u.ac.jp},  
          Masanori H{\sc anada}$^{b}$\footnote
           {
E-mail address : masanori.hanada@weizmann.ac.il},   
          Tomoyoshi H{\sc irata}$^{a}$\footnote
           {
E-mail address : hirata@gauge.scphys.kyoto-u.ac.jp} and 
Hidehiko S{\sc himada}$^{c}$\footnote
           {
E-mail address : Hidehiko.Shimada@aei.mpg.de}

\vspace{0.3cm}

$^a$           
{\it Department of Physics, Kyoto University\\
Kyoto 606-8502, Japan}\\

$^b$           
{\it Department of Particle Physics, Weizmann Institute of Science\\
     Rehovot 76100, Israel }\\

$^c$
{\it
Max-Planck-Institut f\"{u}r Gravitationsphysik, Albert-Einstein-Institut\\
 Am M\"{u}hlenberg 1, D-14476 Potsdam, Germany

}

\end{center}

\vspace{1.5cm}

\begin{center}
  {\bf abstract}
\end{center}

As is well known, coordinates of D-branes are described by 
$N\times N$ matrices. 
From generic non-commuting matrices,
it is difficult to extract
physics, for example, 
the shape of the distribution of positions of D-branes.
To overcome this problem,
we generalize and elaborate on a simple prescription,
first introduced by Hotta, Nishimura and Tsuchiya,
which determines the most appropriate gauge to make the separation
between diagonal components (D-brane positions)
and off-diagonal components.
This prescription makes it possible to extract the
distribution of D-branes directly from matrices.
We verify the power of it by applying it to 
Monte-Carlo simulations for various lower dimensional Yang-Mills
matrix models. In particular, we detect the topology
change of the D-brane bound state 
for a phase transition of a matrix model;  
the existence of this phase transition is expected from
the gauge/gravity duality, and the pattern of the topology change is 
strikingly similar to the counterpart in the gravity side,
the black hole/black string transition.
We also propose a criterion, based on the behavior of
the off-diagonal components, which determines when our prescription
gives a sensible definition of D-brane positions.
We provide numerical evidence that 
our criterion is satisfied for the typical distance between D-branes.
For a supersymmetric model, positions of D-branes can be defined even at a shorter
distance scale.
The behavior of off-diagonal elements found in this analysis
gives some support for previous studies of D-brane bound states.

\newpage 
\section{Introduction}
\hspace{0.51cm}

The most characteristic property of D-branes 
is that each coordinate of $N$ D-branes is promoted to an $N\times N$
matrix $X^\mu$, rather than a set of $N$ numbers~\cite{Witten95}.  
One usually identifies the $N$ diagonal components as 
the positions of D-branes in the conventional sense,
and the $N(N-1)$ off-diagonal components
as degrees of freedom of open strings connecting them.
However, these $N^2$ degrees of freedom mix under the $U(N)$ gauge symmetry
of the system of D-branes, and
therefore there is inherent indeterminacy in this identification 
of the positions of D-branes.

However, of course, one should not totally abandon the concept of 
positions for D-branes. For example, when D-branes are
``far apart", there should be a gauge where
distances between $N$ points specified by diagonal elements 
are large, and then off-diagonal elements should be dynamically suppressed
by the mass-term induced from the potential term,
\begin{equation}
- Tr [X^\mu, X^\nu]^2
=
\sum_{\mu\neq\nu}\sum_{i < j}(X^\mu_{ii}-X^\mu_{jj})^2|X_{ij}^\nu|^2
+
O(({\rm off\ diagonal})^3).  
\end{equation}
Although the $U(N)$ symmetry allows us to make arbitrary
unitary transformations to the almost diagonal matrices described above,
it is clear that 
it makes much more sense 
to call diagonal elements in the gauge where off-diagonal elements 
are very small as positions of D-branes, rather than those in
the general gauge.

In more general situations it becomes more obscure
whether a similar concept of positions of D-branes exists.
However, for many applications, it would be useful
if one could define the concept of positions for D-branes 
(in terms of $N$ numbers) 
for general situations, since it is much easier to grasp intuitively.
One application we have in mind is the gauge/gravity correspondence 
\cite{Maldacena97,IMSY98}:
the black hole geometry on the gravity side should be
related to the shape of the D-brane bound state on the Yang-Mills side,
and in order to study the latter it will be essential to define the
positions of D-branes in an appropriate way.

In this paper we use a simple prescription,
introduced in \cite{HNT98} for a slightly different purpose, 
to define positions of D-branes. 
The essential idea is to maximize the diagonal elements,
in an appropriate measure,
by an unitary transformation. 
For example, for the zero-dimensional matrix model
(the bosonic analogue of
an effective action for D$(-1)$ branes)
\begin{eqnarray}
S=-\frac{N}{4}\ Tr[X_\mu,X_\nu]^2,
\label{action:bosonicIKKT}
\end{eqnarray}
where $X_\mu\ (\mu=1,\cdots,D)$ are $N\times N$ Hermitian matrices,
we perform the $U(N)$ transformation such that the quantity
\begin{eqnarray}
\sum_{\mu=1}^D
\sum_{i=1}^N
\left[\left(
UX_\mu U^\dagger
\right)_{ii}\right]^2
\end{eqnarray}
takes the maximum value. 
It is obvious that diagonal components
in this gauge coincide with the D-brane positions
discussed above for the (almost) mutually diagonalizable case.  

We analyze, using the above prescription (and its generalizations)
for determination of D-brane positions,
the results of Monte-Carlo simulations for various zero- and one-dimensional 
Yang-Mills matrix models\footnote{
Yang-Mills theories are effective theories for the low-energy dynamics of 
D-branes. For high energy processes one should also incorporate the higher excitation modes 
of open strings. Ideas discussed in this paper would be also useful 
even if these higher modes are included. }. 
Our aim is twofold: the first is to verify the power
of this prescription. 
In particular, we will see that the method allows us
to detect the topology change of the distribution
of D-branes; this is consistent with the
expectation from the gauge/gravity correspondence,
and corresponds to the transition between
black strings and black holes.~\footnote{
The phase transition itself was identified and studied
previously, but it has not been shown
that this phase transition is accompanied by the
change in the shape (including information in transverse directions)
of the D-brane bound state.}
The second is to give a criterion to 
determine when the use of this prescription
gives rise to a sensible definition of positions of D-branes, in other words,
to give at least a partial answer to the important question,
``when are the D-brane positions well-defined?"
The basic idea is to see whether off-diagonal elements
are governed by simple Gaussian distributions
which are determined by diagonal elements;
in such a case one can regard dynamics of off-diagonal elements
as sub-dominant, compared to that of diagonal elements,
justifying the separation between diagonal and off-diagonal elements.
We find that this criterion is satisfied with a relatively 
short length scale, compared to the size of the whole D-brane distribution,
which validates the use of our method to examine the shape of
the D-brane bound state. 

This paper is organized as follows.
In \S~\ref{sec: gravity dual} 
we introduce the matrix models we consider,
and review what the gauge/gravity 
correspondence implies for 
qualitative behaviors of the models.
These models are simplified versions
of maximally supersymmetric Yang-Mills theories 
for D-branes; we also explain 
the relevance of the simplified models for our purpose.  
In \S~\ref{sec:basic idea} we explain our main ideas
using the zero-dimensional matrix model as an example:
we explain in detail the maximal diagonalization procedure introduced above
and our criterion to determine when this procedure provides
a sensible definition of D-brane positions.
We also show the result of the maximal diagonalization for the standard
example, the fuzzy sphere. 
The results of Monte-Carlo simulations
for the diagonal and the off-diagonal elements 
are respectively given in \S~\ref{sec:diag distribution} 
and \S~\ref{sec:offdiag distribution}. 
In \S~\ref{sec:MQM} we study the bosonic matrix quantum mechanics. 
We first clarify its phase structure 
in \S~\ref{sec:MQM phase structure}.
In \S~\ref{sec:diagonalization procedure for MQM} we make an
appropriate generalization of
the maximal diagonalization to non-zero spacetime dimensions
by combining it with the T-duality transformations for 
the Yang-Mills theory~\cite{Taylor96}. In \S~\ref{sec:BH/BS transition}
we apply the method to the Monte-Carlo data:
we read off the topology change in the shape of the D-brane bound state,
a counterpart of the black hole/black string transition. 
In \S~\ref{sec:SUSY} we show preliminary 
results for a supersymmetric matrix model;
in this case we find that our criterion is satisfied 
even at shorter distances between D-branes. 
We also discuss the implication for 
the structure of the ground state. 
\S~\ref{sec:discussion} is devoted to 
conclusion and discussions.

\section{Dual gravity description and black hole/black string transition}
\label{sec: gravity dual}
\hspace{0.51cm}
In this section we introduce a class of matrix models
in lower dimensional spacetime, which will be considered in this paper.
We also review the gauge/gravity correspondence \cite{Maldacena97,IMSY98} 
which will give us 
some expectations about qualitative behaviors of these models,
in particular phase transitions corresponding to
the transition between the black hole (BH) and the black string (BS)
on the gravity side~\cite{GL93, AMMW04,Su,BKR,MaSa,MaSa2,HO}.
The presentation here basically follows \cite{AMMW04}. 
For review articles, see \cite{GregoryLaflammeReview}.
A world-sheet approach is proposed
in \cite{KS07}, to understand this duality.

The models we consider are simplified versions
of the models for D$p$-branes at finite temperature.
Low energy dynamics of $N$ D$p$-branes is described by 
supersymmetric Yang-Mills (SYM) in $(1+p)$-dimension with the $U(N)$ gauge group, 
which can be obtained from the ten-dimensional ${\cal N}=1$ SYM 
through the dimensional reduction~\cite{Witten95}. 
By Euclideanizing the time coordinate and by compactifying it with a period $\beta$, 
we obtain the finite temperature theory with the temperature $1/\beta$.

Let us start with the maximally supersymmetric $U(N)$ SYM
theory in $(1+1)$ dimension at finite temperature, 
\begin{eqnarray}
S_{2dSYM}
=
\frac{N}{4\lambda}
\int_0^{1/T_H}dt
\int_0^L dx\ 
Tr\left(
F_{\mu\nu}^2
+
2(D_\mu X_I)^2
-
[X_I,X_J]^2
+
{\rm (fermion)}
\right), 
\label{RF2DSYM}
\end{eqnarray}
where the temporal direction $t$ and 
spatial directions $x$ are compactified with length $1/T_H$ and $L$, respectively. 
Our notations are as follows: 
$T_H$ is the temperature (the subscript $H$ means it is identified with 
the Hawking temperature of the dual black brane), $\lambda$ is the 't Hooft coupling, 
$X_I\ (I=1,\cdots,8)$ are adjoint scalars. 
For fermionic fields the anti-periodic boundary condition
should be taken along the temporal direction. 
 
In the high temperature limit of SYM, $T_H\to\infty$, 
only zero-modes in the temporal direction survive;
in particular,
all fermionic fields decouple. 
Then we obtain the bosonic matrix quantum mechanics
\begin{eqnarray}
S_{1d,bos}
=
\frac{N}{\lambda T_H}
\int_0^L dx\ 
Tr\left(
\frac{1}{2}(D_x X^\prime_P)^2
-
\frac{1}{4}
[X^\prime_P,X^\prime_Q]^2
\right), 
\label{ActionBosonicMQMWithoutAbsorption}
\end{eqnarray}
where $P=(t,I)$, $X^\prime_I=X_I$, and $X^\prime_t$ is the adjoint scalar field
coming from the $t$-component of the gauge field. 
We will study models of this type in \S~\ref{sec:MQM}. 
We remark that it is far from clear whether one can use the gauge/gravity duality 
for this high temperature limit; for the high temperature limit of 
a D-brane bound state, effects of the higher excitation modes 
of open strings are not negligible. 
The reason for studying these high-temperature Yang-Mills (YM) theories
such as (\ref{ActionBosonicMQMWithoutAbsorption}) 
is that they are very useful to test our ideas because 
(i) Monte-Carlo simulations are much more tractable for them, and 
(ii) one can gain insight into their properties, in particular their phase structures,
from corresponding low temperature theories (where one can use the gauge/gravity correspondence).
These models are also interesting in their own rights. 

In the low temperature region~\footnote{
The dimensionless quantity 
$\lambda/T_H^2$ 
can be considered as an effective coupling constant,
and it is customary to call 
the low temperature region also as the strong coupling region.  
}, 
it is conjectured in \cite{Maldacena97, IMSY98} that SYM
theory (\ref{RF2DSYM}) is dual to type IIB superstring theory
on a certain background. The background is obtained
by taking the near-horizon limit of the
near-extremal black 1-brane solution in type IIB supergravity 
compactified on a circle. 
The solution in the string frame is \cite{HS91,IMSY98}
\begin{eqnarray}
& &
ds^2
=
\alpha^\prime\left\{
\frac{U^3}{\sqrt{\lambda d_1}}
\left[
-\left(
1-\frac{U_0^6}{U^6}
\right)
dt^2
+
dx^2
\right]
+
\frac{\sqrt{\lambda d_1}}{U^3\left(1-\frac{U_0^6}{U^6}\right)}
dU^2
+
\sqrt{\lambda d_1}U^{-1}d\Omega_7^2
\right\}, 
\nonumber\\
& &
e^\phi
=
\frac{2\pi\lambda}{N}
\sqrt{\frac{\lambda d_1}{U^6}}, 
\label{black 1 brane}
\end{eqnarray}
where 
$d_1
=
2^6\pi^3$,   
and $0\le x<L$ parametrizes the $S^1$ direction. 
Thermodynamic quantities should be calculated in the Einstein frame 
defined by 
$G_{\mu\nu}^{Einstein}=e^{-\phi/2}G_{\mu\nu}^{string}$. 
The Hawking temperature is 
\begin{eqnarray}
T_H
=
\frac{6U_0^2}{4\pi\sqrt{\lambda d_1}}, 
\label{temperature_and_radius} 
\end{eqnarray}
and the ADM energy and the entropy are respectively given by 
\begin{eqnarray}
E
&=&
\frac{1}{2^7\cdot 3\pi^5\lambda^2}
N^2U_0^6L, 
\\
S
&=&
\frac{1}{2^4\cdot 3\pi^{5/2}\lambda^{3/2}}
N^2U_0^4L. 
\end{eqnarray}
In order for the supergravity description to be valid, 
the string excitation and winding modes  
associated to the
$x$-direction should be much heavier than the KK modes 
along $S^7$, that is, 
\begin{eqnarray}
\frac{1}{\sqrt{\lambda}L^2}\ll T_H\ll\sqrt{\lambda}. 
\end{eqnarray}
Note that string loop corrections are negligible, because  
the string coupling $e^\phi$  is vanishingly small 
in the planar limit,
where 
$N$ is taken to be large while
$\lambda$ and $T_H$ are finite.
 
By applying the T-duality transformation along the
$x$-direction, we obtain the black string solution\footnote{
In this paper, we call this solution as the black string. 
We do not call the solution to the IIB supergravity (\ref{black 1 brane}) 
as the black string. 
},
which is also called the smeared $D0$-brane solution,
in type IIA supergravity, 
\begin{eqnarray}
& &
ds^2
=
\alpha^\prime
\left\{
-\frac{U^3}{\sqrt{\lambda d_1}}
\left(
1-\frac{U_0^6}{U^6}
\right)
dt^2
+
\frac{\sqrt{\lambda d_1}}{U^3}
\left[
\frac{dU^2}{\left(1-\frac{U_0^6}{U^6}\right)}
+
d\tilde{x}^2
\right]
+
\sqrt{\lambda d_1}U^{-1}d\Omega_7^2
\right\}, \label{T-dualized metric}
\\
& &
e^\phi
=
(2\pi)^2\frac{\lambda}{N}
\left(
\frac{\lambda d_1}{U^6}
\right)^{3/4}
\frac{1}{L}\ , 
\end{eqnarray}
where $\tilde{x}$ varies from $0$ to $(2\pi)^2/L$. 
Again the string coupling $e^\phi$ is vanishingly small 
in the planar limit. 
Thermodynamic quantities like
the Hawking temperature, the ADM energy and the entropy 
are unchanged by the T-duality transformation. 
This type IIA supergravity description 
is valid in a different parameter region; 
the winding modes are now negligible when 
\begin{eqnarray}
T_H\ll\frac{1}{L}.  
\label{IIA_validity_1}
\end{eqnarray}
The $\alpha^\prime$ corrections are negligible when 
\begin{eqnarray}
T_H\ll\sqrt{\lambda},
\label{IIA_validity_2}
\end{eqnarray}
which is the same condition as before.  

In this solution $D0$-branes are smeared and form BS
winding on the compact direction. The essence of the 
BS/BH transition is to compare the free energy of
BS with that of a BH solution with the same charge 
and temperature. In the BH solution, $D0$-branes clump to  
a small region.
Such a solution in compact space
can be approximated very well 
by the black 0-brane solution to IIA supergravity 
in noncompact space. (See \cite{KW04}, for example.)
The solution is \cite{HS91,IMSY98}
\begin{eqnarray}
& &
ds^2
=
\alpha^\prime\left\{
-\frac{U^{7/2}}{\sqrt{\lambda d_0}}
\left(
1-\frac{U_0^7}{U^7}
\right)
dt^2
+
\frac{\sqrt{\lambda d_0}}{U^{7/2}\left(1-\frac{U_0^7}{U^7}\right)}
dU^2
+
\sqrt{\lambda d_0}U^{-3/2}d\Omega_8^2
\right\}, \nonumber
\\
& &
e^\phi
=
\frac{(2\pi)^2\lambda}{N}
\left(\frac{\lambda d_0}{U^7}\right)^{3/4},
\qquad
T_H=\frac{7U_0^{5/2}}{4\pi\sqrt{d_0\lambda}},  
\label{BH metric}
\end{eqnarray}
where 
$d_0
=
2^7\pi^{9/2}\Gamma(7/2)$. This solution is conjectured to be dual to a 
$(1+0)$-dimensional SYM,
and   
$\lambda,\ T_H$ are the 't Hooft coupling and the temperature 
for the dual theory.~\footnote{
Recently this duality has been confirmed very precisely 
by studying the strong coupling regime directly using the Monte-Carlo simulation 
\cite{AHNT07,CW08,HMNT08,HHNT08}. 
The expectation value of the Wilson loop \cite{HMNT08} and the internal energy 
\cite{HHNT08} are consistent with 
their counterparts on the gravity side 
including $\alpha'$-corrections.    
An interesting approach for understanding the duality has been proposed 
in \cite{Smilga08}. 
} 
The Hawking temperature is unchanged by the T-duality transformation
and $\lambda$ is related to that of $(1+1)$-dimensional SYM by 
\begin{eqnarray}
\lambda_{(1+0)}=\frac{\lambda_{(1+1)}}{L}.
\end{eqnarray} 
The action of the (0+1)-dimensional SYM is
\begin{eqnarray}
S_{1dSYM}
=
\frac{N}{\lambda}
\int_0^{1/T_H}dt\  
Tr\left(
\frac{1}{2}(D_t X_I)^2
-
\frac{1}{4}
[X_I,X_J]^2
+
{\rm (fermion)}
\right).   
\label{1dSYMaction}
\end{eqnarray}
We will study an analogous model in \S~\ref{sec:SUSY}. 
In the high temperature regime 
the action (\ref{1dSYMaction}) reduces to the bosonic matrix model 
(\ref{action:bosonicIKKT}) with $D=10$. 
We study $D=3$ and $D=4$ cases in \S~\ref{sec:0dMM}. 
 
The BH/BS transition point can be identified by
a careful comparison of the free energy  
of the two solutions. 
The critical temperature is 
\begin{eqnarray}
T_c\sim\frac{1}{L^2\sqrt{\lambda}},  
\end{eqnarray} 
below which BS breaks down to BH. 
Here $\lambda$ is the 't Hooft coupling constant in $(1+1)$-dimension. 
Note that not type IIB but type IIA supergravity is valid at $T_c$, 
comparing with equations (\ref{IIA_validity_1}) and (\ref{IIA_validity_2}),
validating our description of the transition using IIA supergravity.
If we fix the temperature $T_H$ and vary $L$, then 
BS appears at large $L$. 
This is not counterintuitive,
because if $L$ is large then the compactification radius in the T-dualized 
picture is small; it is natural that BH ceases to exist
if the compactification radius becomes smaller -- eventually it winds on 
the compactified dimension and forms BS. 

The gauge theory dual of this transition was previously studied
using the Wilson line operator $W={\rm P}\exp\left(i\int_0^{L}dx A_x\right)$,
winding on the compactified direction,
as the order parameter.  
The basic idea is as follows.
As $W$ is unitary, its eigenvalues lay on a unit circle on
the complex plane. They correspond to 
the coordinates of $D0$-branes along the compactified direction. 
In the BS phase, the eigenvalues
should be distributed uniformly on the
unit circle and hence the trace of
the Wilson line should be zero. On the other hand, 
in the BH phase the eigenvalues are clumped in a small region 
on the unit circle and hence the trace should be non-zero. 
Thus, the transition from the BS phase to the BH phase
can be identified with the breakdown 
of the global $U(1)$ symmetry, 
$A_x\to A_x+ \mbox{const.}$, which implies $W\to e^{i\theta}W$. 
On the gauge theory side, 
the breakdown of the $U(1)$ symmetry has not yet been studied 
for the low temperature regime, where the gauge/gravity 
correspondence should be valid.
In \cite{AMMW04} the bosonic matrix quantum mechanics has been studied 
as the high-temperature limit of the 
$(1+1)$-dimensional model,
and the transition, precisely characterized by the
breakdown of the $U(1)$ symmetry, is found. 
See \cite{JW00} for an earlier work. 
It is natural to consider this transition as
the continuation of the transition in the low-temperature regime
predicted by the gauge/gravity correspondence.
In \cite{KNT07} this transition has been studied further in detail 
and it turned out that there are actually two successive phase transitions. 
They can be characterized as ``uniform black string to non-uniform black string" 
and ``non-uniform black string to black hole" transitions; 
here (non-)uniformity refers to that in the compactified direction.
At low temperature, 
supergravity analysis suggests there is only one transition, 
from uniform black string to black hole. 
It is an open problem to clarify the
structure of the phase transitions at 
intermediate temperature. 
  
Monte-Carlo simulations for 
the maximally supersymmetric matrix models associated with D-branes
would allow us to 
compare the results 
with the gravity dual. 
In this paper 
we shall only consider simplified versions of matrix models for D-branes, 
with fewer scalar and fermionic fields.    
The reason for this is that
the simplified models 
are suitable for testing our ideas about defining positions of D-branes,
because
the simplified models are qualitatively 
similar to the full matrix models for D-branes, 
and Monte-Carlo simulations are more  
tractable for the simplified versions\footnote{
The technical obstacles for simulating the full matrix models for D-branes 
are as follows.
First, Monte-Carlo simulations for the full model is much more 
computationally heavy. 
This problem is severe especially for supersymmetric models, 
because of the necessity of the calculation of the fermionic determinant.   
In the case of the bosonic models, 
the simulation cost itself does not depend so severely on 
the number of the matrices $D$. 
However, we expect that the convergence to the large-$N$ distribution is slower at larger $D$,  
because the diagonal components are distributed in ${\mathbb R}^D$, 
and hence typically we have only $N^{1/D}$ meshes for each direction. 
For example for D$0$-branes, say,  we have $D=9$, and
if we want to divide each directions into four parts,
then $N$ should be $4^{9}$, which is hopelessly large.
Second, supersymmetric matrix models with 16 supercharges 
have the notorious sign problem 
due to the pfaffian of the Dirac operator being complex. 
The sign problem potentially makes 
the construction of the distribution of the diagonal components 
very difficult. 
The reason is as follows.
The notion of the distribution of the diagonal components  fundamentally 
relies on the expectation that physics of the matrix model can be studied 
by considering typical samples.
Usually typical samples are easily obtained by Monte-Carlo methods. 
However, if the sign problem is present,
it can be difficult to obtain these typical samples,
because there might be many spurious configurations 
whose effects simply cancel out.  
The sign problem does not exist for
supersymmetric matrix models with four supercharges~\cite{KNS98,AABHN00}.
}. 
For example, as we will see in \S~\ref{sec:MQM}, the bosonic matrix quantum mechanics with 
fewer scalar fields has the same phase structure as the one with 9 scalars \cite{KNT07}, 
which is the high-temperature limit of the D1-brane matrix model. 
Also, simulation results for 4-supercharge \cite{HNT07,CW07} and 
16-supercharge \cite{AHNT07,CW08} matrix models are qualitatively the same.

\section{Matrix model in zero dimension and black hole geometry}
\label{sec:0dMM}
\hspace{0.51cm}
In this section we analyze 
the bosonic zero-dimensional matrix model (\ref{action:bosonicIKKT}). 
We first explain two main ideas
in our paper: the maximal diagonalization procedure,
first introduced by \cite{HNT98},
and our criterion which determines when the concept of  
D-brane positions (in terms of $N$ numbers, not $N\times N$ matrices) makes sense.
We then analyze results of  Monte-Carlo simulations according to these ideas. 
\subsection{Basic ideas}
\label{sec:basic idea}
\hspace{0.51cm}
As explained in the previous section, 
at high temperature, supersymmetric matrix quantum mechanics 
describing D0-branes (\ref{1dSYMaction}) reduces to the bosonic matrix model 
(\ref{action:bosonicIKKT}) with $D=10$.  
This model with smaller values of $D$ 
has been studied extensively in \cite{HNT98}.   
The action is 
\begin{eqnarray}
S
=
-\frac{N}{4}
\sum_{\mu,\nu=1}^D
Tr [X_\mu,X_\nu]^2, \label{Action0DMMLater}
\end{eqnarray}
where $\mu,\nu=1,\cdots,D$ and $X_\mu$'s are traceless Hermitian matrices. 
 
First, let us introduce the maximal diagonalization procedure~\cite{HNT98}.
According to the standard interpretation, diagonal elements of $X$'s 
correspond to positions of D-branes and off-diagonal elements correspond
to degrees of freedom of open strings 
connecting two different D-branes \cite{Witten95}. 
But of course ``diagonal" and ``off-diagonal" 
are gauge-dependent notions -- 
in order to obtain a unique definition of D-brane positions, 
we should choose an appropriate gauge. 

Let us consider the case where all branes are well separated.
In this case, open strings are long and heavy, and hence 
not excited easily.  
If we neglect these very small excitations,
we can simultaneously diagonalize the matrices describing the branes,
and it is clear that one can identify 
the diagonal elements in this gauge as the 
positions of D-branes.
Now let us gradually decrease the distances between the branes.
Then we come to a regime where open string excitations are not negligible,
but are still very small compared to typical values 
of diagonal elements. In this case, although in general gauges
off-diagonal and diagonal elements are of the same order,
in some gauges diagonal elements
are much larger than the off-diagonal elements.
It is still at least natural to call diagonal elements in these special gauges
as positions of D-branes.
This consideration leads us to {\it maximally diagonalize} 
matrices using an unitary matrix $U=U_{max}\in U(N)$, 
where we maximize the following quantity,  
\begin{eqnarray}
\sum_{\mu=1}^D
\sum_{i=1}^N
\left[\left(
UX_\mu U^\dagger
\right)_{ii}\right]^2. 
\label{measure for diagonalization}
\end{eqnarray} 
Clearly, if $X_\mu$'s are simultaneously diagonalizable, 
$U_{\max}X_\mu U_{\max}^\dagger$ are diagonal. 
In general, $U_{\max}X_\mu U_{\max}^\dagger$ are as close to 
simultaneously diagonalizable as possible. 
It is reasonable to define positions of D-branes as
diagonal elements in this gauge,   
$\left(
U_{max}X_\mu U_{max}^\dagger
\right)_{ii}$. 
Note this notion of positions is gauge-independent, 
as  gauge-equivalent sets of matrices end up with the same diagonal components
after the maximal diagonalization, if we neglect the unlikely case
where there are several minima
for the quantity (\ref{measure for diagonalization}).

We wish to note that the measure for the maximal diagonality 
we employ, (\ref{measure for diagonalization}),
although natural, is certainly not the unique choice.
One might also try to maximize  
$\sum_{\mu=1}^D
\sum_{i=1}^N
\left|\left(
UX_\mu U^\dagger
\right)_{ii}\right|$, for example. 
An advantage of our measure (\ref{measure for diagonalization})
is that it respects the $SO(D)$ rotational symmetry
of the model.
Another important property of (\ref{measure for diagonalization}) is that 
maximizing the diagonal elements amounts at the same time
to minimizing the off-diagonal elements,
because the matrix norm,
\begin{eqnarray}
\sum_{\mu=1}^D
\sum_{i,j=1}^N
\left|\left(
UX_\mu U^\dagger
\right)_{ij}\right|^2
=
\sum_{\mu=1}^D
\sum_{i=1}^N
\left|\left(
UX_\mu U^\dagger
\right)_{ii}\right|^2
+
\sum_{\mu=1}^D
\sum_{i\neq j}^N
\left|\left(
UX_\mu U^\dagger
\right)_{ij}\right|^2,
\label{RFMatrixNorm}
\end{eqnarray}
is invariant under unitary transformations. We will revisit this issue 
in \S~\ref{sec:diagonalization procedure for MQM}.

The important next step is to consider 
when this maximal diagonalization procedure 
gives a  sensible definition of positions of D-branes.
As is clear from the above discussion, it is the behavior of
the off-diagonal elements that is crucial for the consideration
of when D-brane positions admit a sensible definition.
Again, when all branes are well separated, certainly
positions of D-branes should be well-defined. 
We wish to stress, however, that off-diagonal elements
are non-zero, although very small, even in this case.
To be precise,
let us remind that the action can be written as 
\begin{eqnarray}
S=
N\sum_{\mu\neq\nu}\sum_{i < j}(X^\mu_{ii}-X^\mu_{jj})^2|X_{ij}^\nu|^2
+
O(({\rm off\ diagonal})^3). \label{SAsSumOfOriginOfGaussianAndHigherOrderTerms}
\end{eqnarray}
When distances between branes $X^\mu_{ii}-X^\mu_{jj}$ are large,
the first term suppresses the excitation of the off-diagonal elements:
because of the Boltzmann factor $e^{-S}$, the distribution of 
off-diagonal elements should become a Gaussian with a very narrow width.
The second term representing the higher order interactions
is negligible because of this narrow width.

The criterion we propose in this paper is motivated by this observation.
We require first that the distribution of off-diagonal elements have 
an almost Gaussian form,
and then compare the width of the Gaussian with that calculated from 
(\ref{SAsSumOfOriginOfGaussianAndHigherOrderTerms}). 
We will say that the positions of D-branes are well-defined 
when the observed width of the Gaussian agrees well with the 
theoretical prediction from (\ref{SAsSumOfOriginOfGaussianAndHigherOrderTerms}).
Put in another way, 
we require that 
the higher-order interaction terms
to be negligible and the off-diagonal elements
is simply governed by the quadratic part of $S$.
The reasoning behind this is that, when this criterion is satisfied,
the dynamics of the off-diagonal elements is trivial
in the sense that it can be inferred from 
the information of the diagonal elements without any difficulty. 
This justifies to make a separation,
which is essential in defining positions of D-branes,
between diagonal elements, 
containing interesting
dynamics, and off-diagonal elements, playing a sub-dominant role in dynamics
of the system.

A more intuitive way of understanding our criterion is to 
think it as the condition when we can use the simple 
``D-brane+open string" picture (neglecting the open string interactions).
If $O(({\rm off\ diagonal})^3)$ terms are negligible,
transverse modes~\footnote{
As is clear from (\ref{SAsSumOfOriginOfGaussianAndHigherOrderTerms}), 
we should only consider transverse modes:
off-diagonal elements in the longitudinal direction do not have mass terms,
as they are gauge degrees of freedom. In other words,
there are no longitudinal oscillations of open strings.
Furthermore, one can show that the condition of the maximal diagonality 
implies that the off-diagonal components in the longitudinal direction
are always zero.},
which are orthogonal to $\vec{X}_{ii}-\vec{X}_{jj}$,  
behave as the harmonic oscillator with mass $\sqrt{N}|\vec{X}_{ii}-\vec{X}_{jj}|$. 
The $(i,j)$-components can naturally be identified with 
open strings stretching between  
two D-branes sitting at $\vec{X}_{ii}$ and $\vec{X}_{jj}$, 
as the mass of each mode is proportional to length of 
the corresponding open string, as it should be. 
If this relation does not hold, this would mean that 
the distance between the branes is so short, 
many strings are excited, and branes and strings interact so strongly  
that the relative positions of the branes cannot be determined precisely. 

We stress that our criterion, 
which we believe to be reasonably well-motivated, is not the only possible
one, and there can be other criteria which are also useful.
We also want to note highly dynamical nature of the problem
for lower dimensional matrix models:
in these models the vacuum expectation value of scalar fields 
cannot be fixed by putting it as boundary conditions. 
Instead, they should be dynamically generated, as emphasized also in
\cite{BFSS96}.

\begin{figure}[tbp]
\begin{center}
\scalebox{0.3}{
\rotatebox{-90}{
\includegraphics[width=20cm,height=30cm]{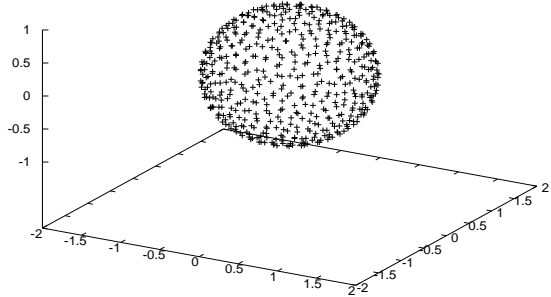}
}}
\caption{Distribution of ``diagonal components" of fuzzy sphere. 
Spin $j=250$. 
}\label{fig:fuzzy sphere spin 250}
\end{center}
\end{figure}

As an illustration of the maximal diagonalization procedure, 
we apply the maximal diagonalization to a standard example, the fuzzy sphere.
We consider three matrices 
\begin{eqnarray}
X_i
=
\frac{1}{\sqrt{j(j+1)}}J_i
\quad
(i=1,2,3), 
\end{eqnarray}
where $J_i$ is $SU(2)$ generator of spin $j=(N-1)/2$.  
We have numerically determined the unitary matrix $U_{max}$ 
which maximizes (\ref{measure for diagonalization}).~\footnote{
In this case the result of the maximal diagonalization is 
not unique;
because of the $SO(3)$ rotational symmetry
of the fuzzy sphere (which can be also realized 
 by a part of unitary transformations) the maximum is degenerate.
Of course, for generic configurations appearing in Monte-Carlo simulations  
there is no such ambiguity.} We find that the
distribution of the diagonal components after the maximal diagonalization 
approaches to a two-sphere of unit radius 
as $j$ becomes large. (See Figure \ref{fig:fuzzy sphere spin 250}.) 
This is the expected result; 
the fuzzy sphere should be a spherical distribution of $D(-1)$ branes.   
We note that typical matrices encountered in Monte-Carlo simulations
are more non-commutative compared to 
the fuzzy sphere. One estimate for the non-commutativity is the ratio
\begin{equation}
\frac{\left(-\frac{1}{N}Tr[X_i,X_j]^2\right)^{1/4}}{\sqrt{\frac{1}{N}TrX_i^2}}
\end{equation}
which is very small when $N$ is large for the non-commutative sphere.
For matrices we encounter in Monte-Carlo simulations,
this ratio is typically of order 
one~\cite{HNT98, BHH08}.
Other methods to extract shape from non-commutative matrices
are proposed in \cite{Hashimoto04} and \cite{Shimada03}.

\subsection{Distribution of the diagonal components -- the shape of D-brane bound state}
\label{sec:diag distribution}
\hspace{0.51cm}
Now we study the bosonic matrix model (\ref{Action0DMMLater}).  
Let us start with the distribution of the diagonal components. 
The distribution is $SO(D)$-symmetric as expected. 
The basic quantity is the radial density function $\rho(r)$, where 
$r=|\vec{X}_{ii}|$, which is normalized so that   
$\int 4\pi r^2\rho(r)dr=1$ for $D=3$ and 
$\int 2\pi^2 r^3\rho(r)dr=1$ for $D=4$. 
To determine $\rho(r)$, we have collected many configurations of matrices 
(for example, 4000 samples for $D=3, N=8$) and evaluated 
the distribution.~\footnote{
In our algorithm, we seek the maximum 
by moving around on the group space, where each step is
a multiplication by a randomly generated $SU(N)$ matrix.
We have checked that two independent diagonalization procedures
give the same result. 
The computational cost increases with $N$,
and for $N=64$ and $96$,
this consistency check sometimes fails
because of the limitation of resources;
we expect the result is at least qualitatively correct.}

\begin{figure}[htbp]
 \begin{minipage}{0.45\hsize}
   \begin{center}
  \rotatebox{-90}{
   \includegraphics[width=50mm]{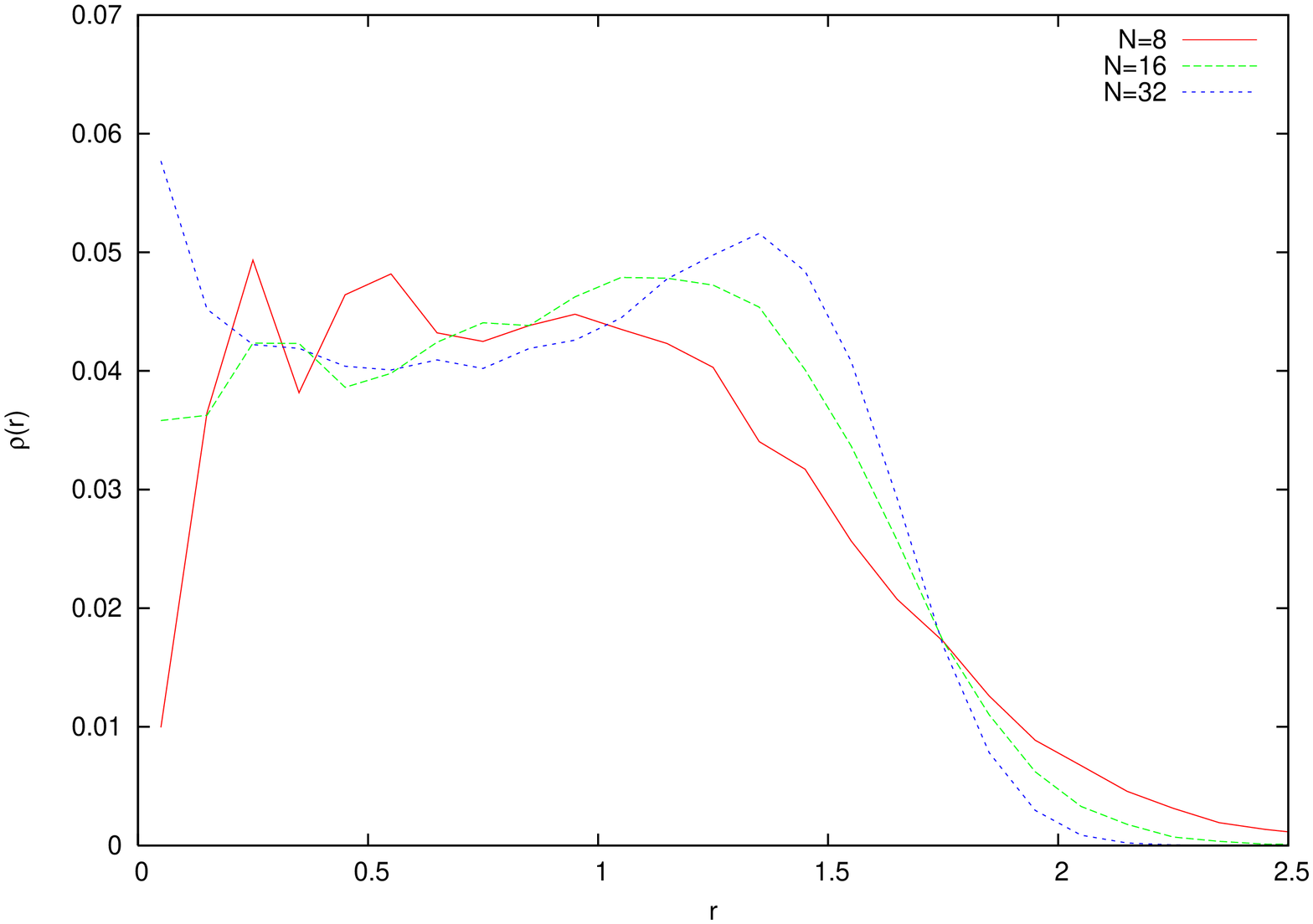}}
  \end{center}
  \caption{Distribution of diagonal components, 
$r=|\vec{X}_{ii}|$ vs $\rho(r)$, in $D=3$ model.  
}
  \label{fig:D3RadDist}
 \end{minipage}
   \begin{minipage}{0.04\hsize}
  \hspace{-5mm}
  \end{minipage}
 \begin{minipage}{0.45\hsize}
  \begin{center}
  \rotatebox{-90}{
   \includegraphics[width=50mm]{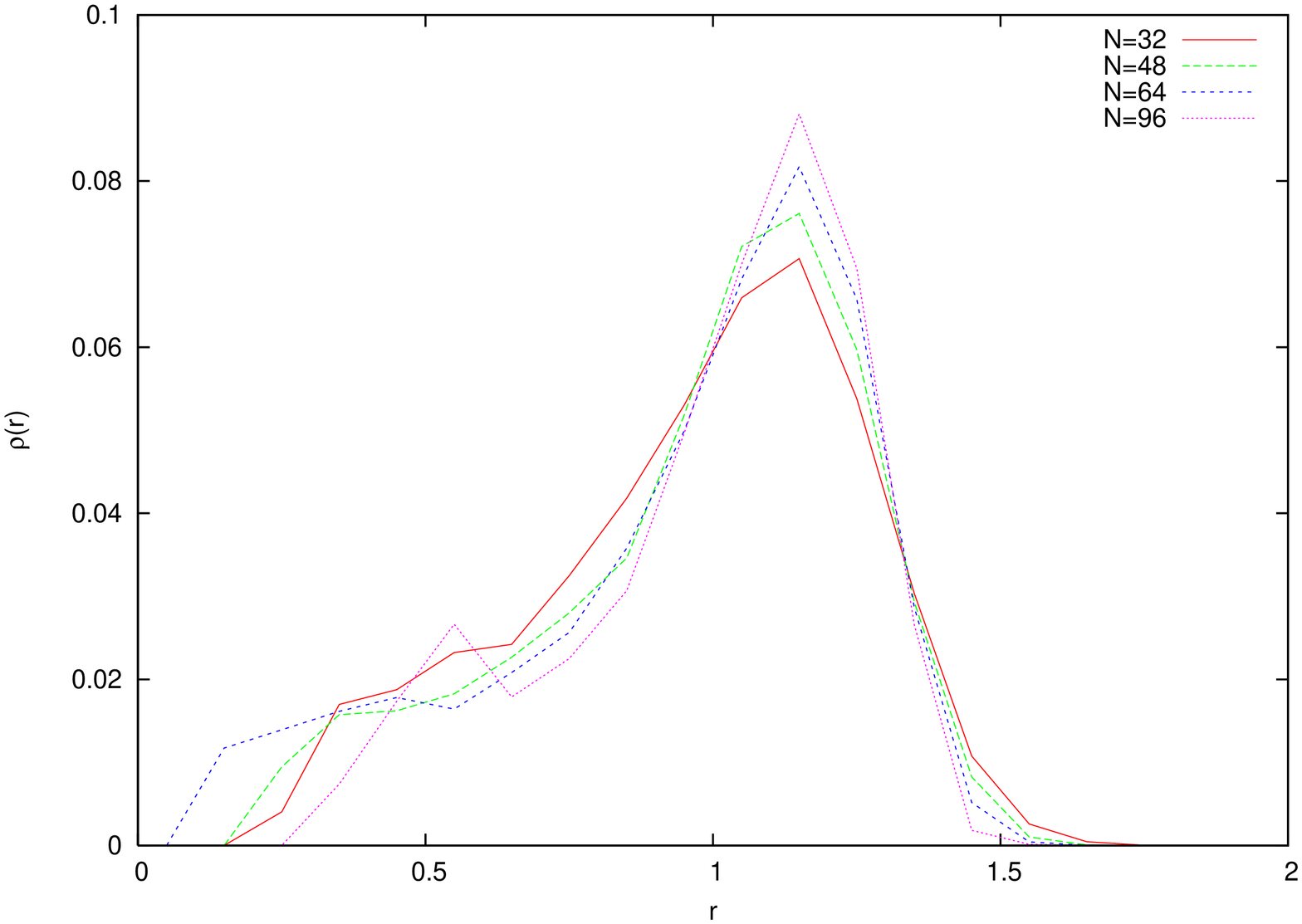}}
  \end{center}
  \caption{$r=|\vec{X}_{ii}|$ vs $\rho(r)$ in $D=4$ model. 
}
  \label{fig:D4RadDist}
 \end{minipage}
\end{figure}

For $D=3$, an almost uniform rigid ball $D^3$
is formed, as we can see from Fig.~\ref{fig:D3RadDist}. 
The unit of length is determined from the normalization 
of the action (\ref{Action0DMMLater}).
Its radius is about $1.7$. 
On the other hand, for $D=4$ the distribution is more like a 
shell, a three-sphere $S^3$ with radius 
$\sim 1.2$, as is shown in Fig.~\ref{fig:D4RadDist}.  
It is not clear whether or not the thickness of the shell
goes to zero at large $N$. We note
that since D-brane positions themselves
have uncertainty as will be discussed in \S~\ref{sec:offdiag distribution},
even if the thickness becomes zero its physical 
implication is not clear.

\subsection{Distribution of the off-diagonal components and validity of the procedure}
\label{sec:offdiag distribution}
\hspace{0.51cm}
In this subsection we study the behavior of off-diagonal components and 
confirm that our criterion discussed in \S~\ref{sec:basic idea}
holds for a wide range of distances between D-branes.

First, we should make a few technical observations regarding
the distribution of off-diagonal components. 
We start from observing that 
the maximally diagonal condition does not fix the gauge completely:
the measure for the diagonality (\ref{measure for diagonalization}) 
is invariant under the 
$U(1)^N$ transformation, generated by $U=diag(e^{i\theta_1},\cdots,e^{i\theta_N})$. 
Under this $U(1)^N$ transformation diagonal components $X_{ii}$ do not change 
while off-diagonal components $X_{ij}$ are multiplied by $e^{i(\theta_i-\theta_j)}$. 
Hence it is natural to study the distribution of the absolute value $|X_{ij}|$. 
The density function for off-diagonal elements
$\rho(|X_{ij}|)$ then acquires the usual volume factor $2\pi |X_{ij}|$ 
in addition to the Gaussian weight.  
Because of the overall factor $N$ in the action, it is also convenient to 
rescale off-diagonal components $X_{ij}$ as $X_{ij}^{new}=\sqrt{N}X_{ij}$, 
so that $y=|X_{ij}^{new}|$ becomes $O(1)$. 

\begin{figure}[htbp]
 \begin{minipage}{0.45\hsize}
  \begin{center}
  \rotatebox{-90}{
   \includegraphics[width=50mm]{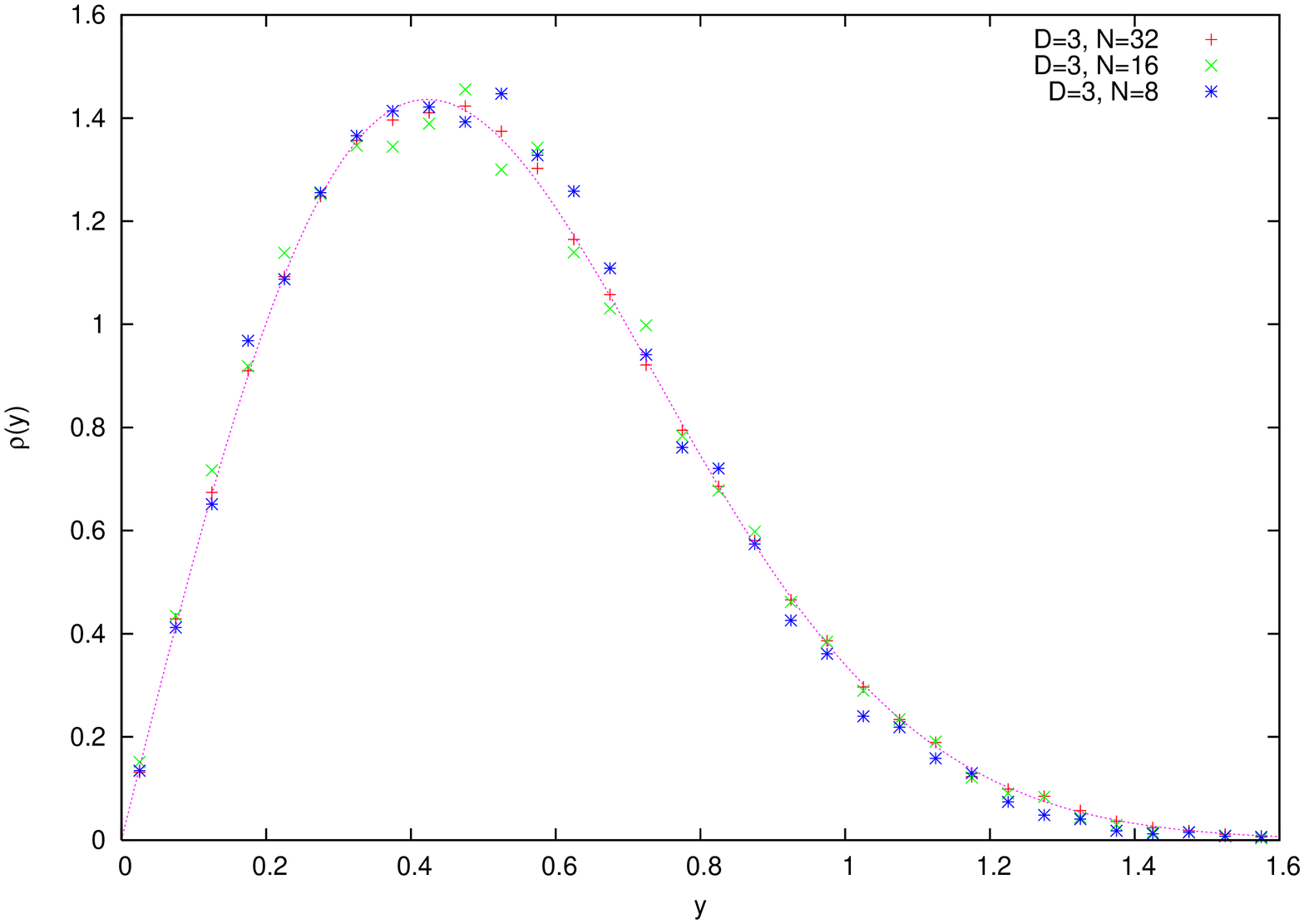}}
  \end{center}
\caption{
The distribution of off-diagonal components $\rho(y)$,  
with $1.50\le d\le 1.55$, in $D=3$ model.   
For $N=32$, the ansatz (\ref{off-diag distribution}) with $a=1.67$ 
provides a reasonable fit.   
}\label{fig:FitD3N32mass150-155}
 \end{minipage}
  \begin{minipage}{0.02\hsize}
  \hspace{-5mm}
  \end{minipage}
 \begin{minipage}{0.45\hsize}
  \begin{center}
  \rotatebox{-90}{
   \includegraphics[width=50mm]{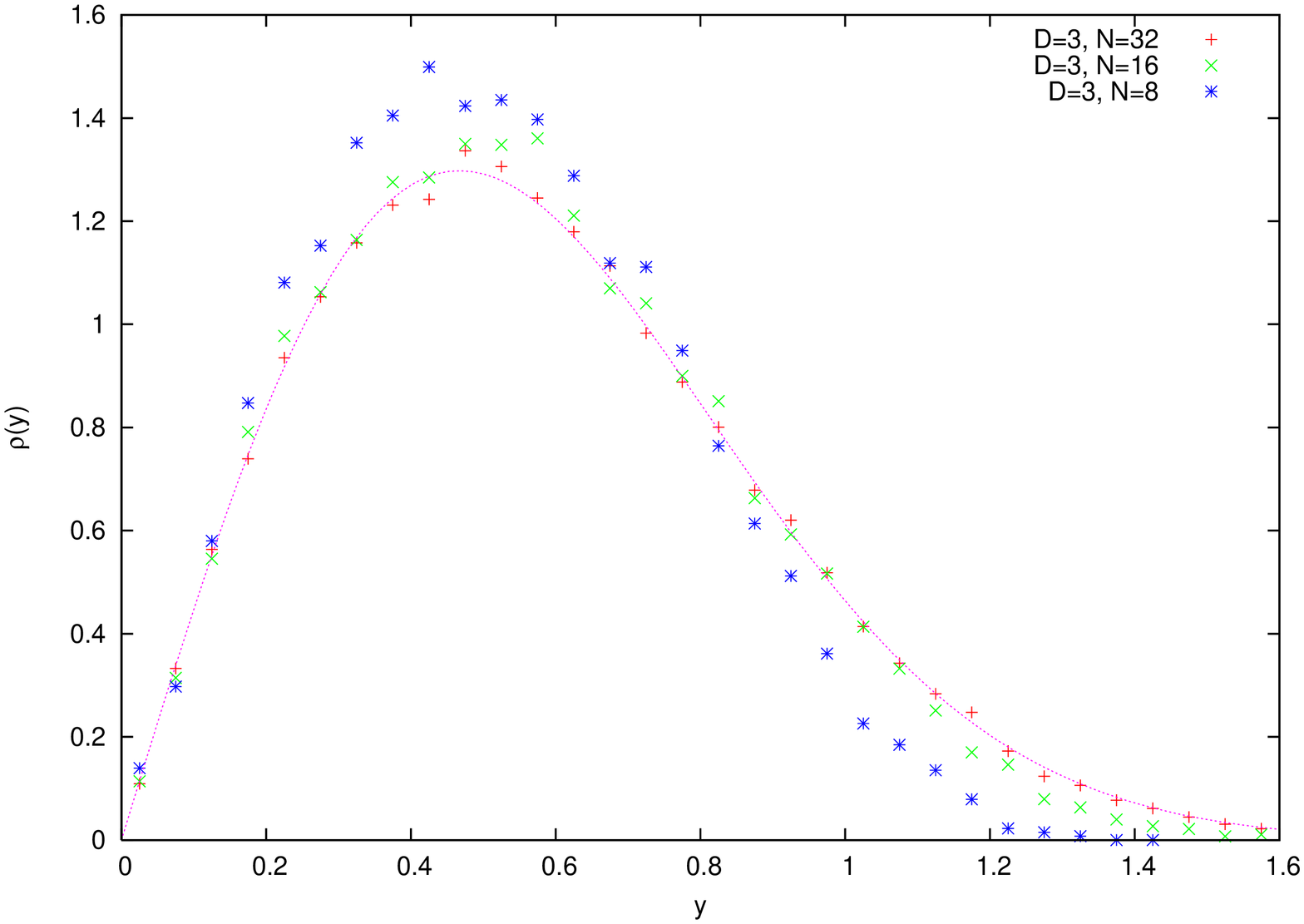}}
  \end{center}
\caption{
The distribution of off-diagonal components $\rho(y)$,  
with $1.00\le d\le 1.05$, in $D=3$ model.   
For $N=32$, the ansatz (\ref{off-diag distribution}) with $a=1.51$ 
provides a reasonable fit. Finite-$N$ corrections are rather large.
}\label{fig:FitD3N32mass100-105}
 \end{minipage}
\end{figure}

If we neglect the interaction terms in 
(\ref{SAsSumOfOriginOfGaussianAndHigherOrderTerms}) according to our criterion, 
the statistical distribution of $y$ 
for fixed value of $\vec{X}_{ii}-\vec{X}_{jj}$
should behave as 
\begin{eqnarray}
\rho(y)
=
2a^2y\exp\left(
-
a^2y^2
\right),
\label{off-diag distribution} 
\end{eqnarray}
with 
\begin{eqnarray}
a=|\vec{X}_{ii}-\vec{X}_{jj}|\equiv d. 
\label{mass=distance relation}
\end{eqnarray}
The overall constant is fixed by the normalization $\int dx \rho(x)=1$. 
We find that (\ref{off-diag distribution})
provides a 
good fit in a large region of $|\vec{X}_{ii}-\vec{X}_{jj}|$,  
see Fig.~\ref{fig:FitD3N32mass150-155} for example.
In general the value of $a$ differs from $d$ but
the ratio $a/d$ as a function of $d$ is close to $1$ except for small $d$.  
This confirms our expectation that when
D-branes are far apart, the distribution of off-diagonal elements
is the Gaussian following from the first term in 
(\ref{SAsSumOfOriginOfGaussianAndHigherOrderTerms}).
At the short distance scale our criterion breaks down.  
As we will show in \S~\ref{sec:SUSY}, 
in the supersymmetric case our criterion holds 
even for shorter distances. 

\begin{figure}[htbp]
 \begin{minipage}{0.45\hsize}
  \begin{center}
  \rotatebox{-90}{
   \includegraphics[width=50mm]{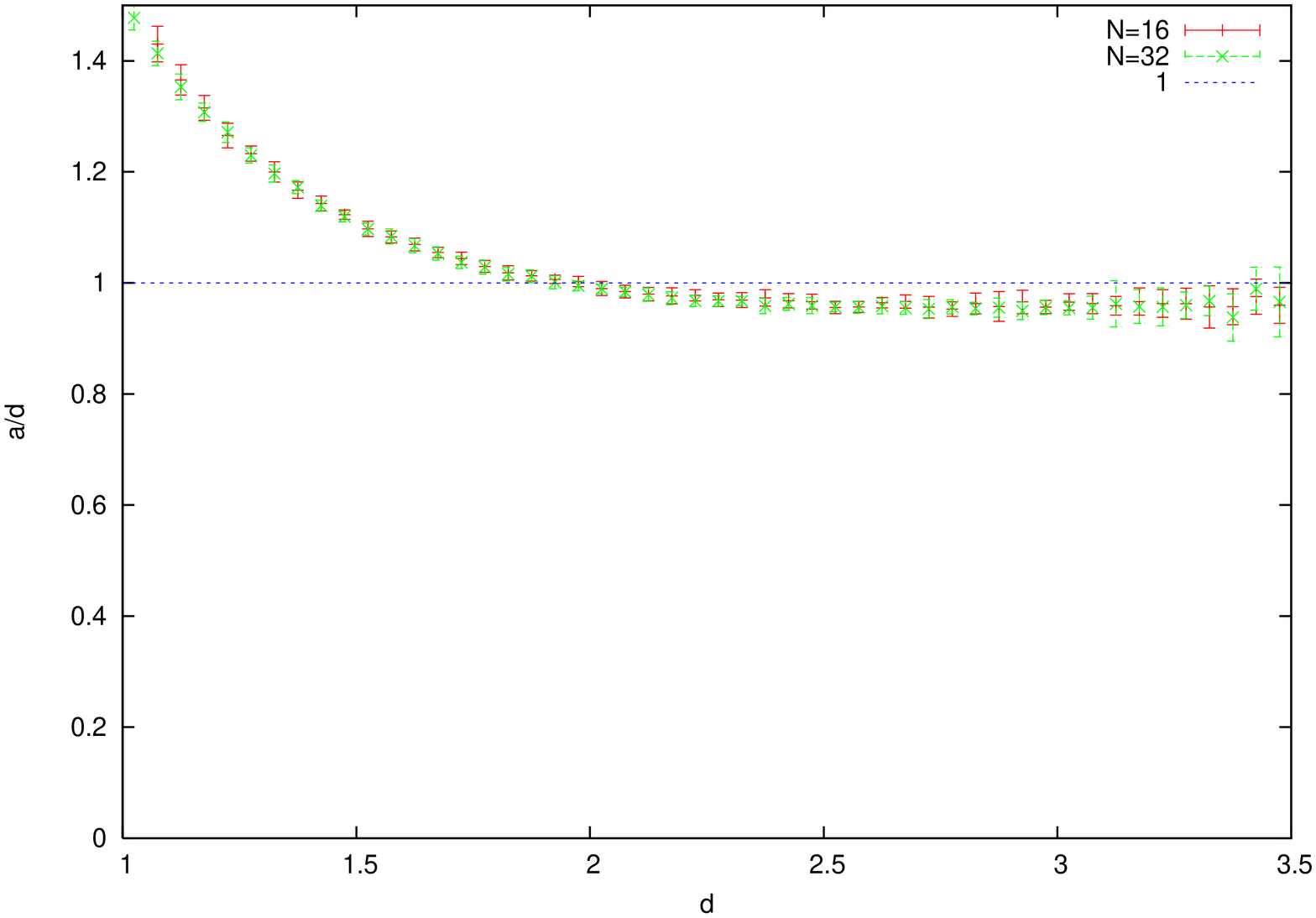}}
  \end{center}
\caption{
$a/d$ vs $d$, $D=3$. 
}\label{fig:Width_D3}
 \end{minipage}
   \begin{minipage}{0.02\hsize}
  \hspace{-5mm}
  \end{minipage}
 \begin{minipage}{0.45\hsize}
  \begin{center}
  \rotatebox{-90}{
   \includegraphics[width=50mm]{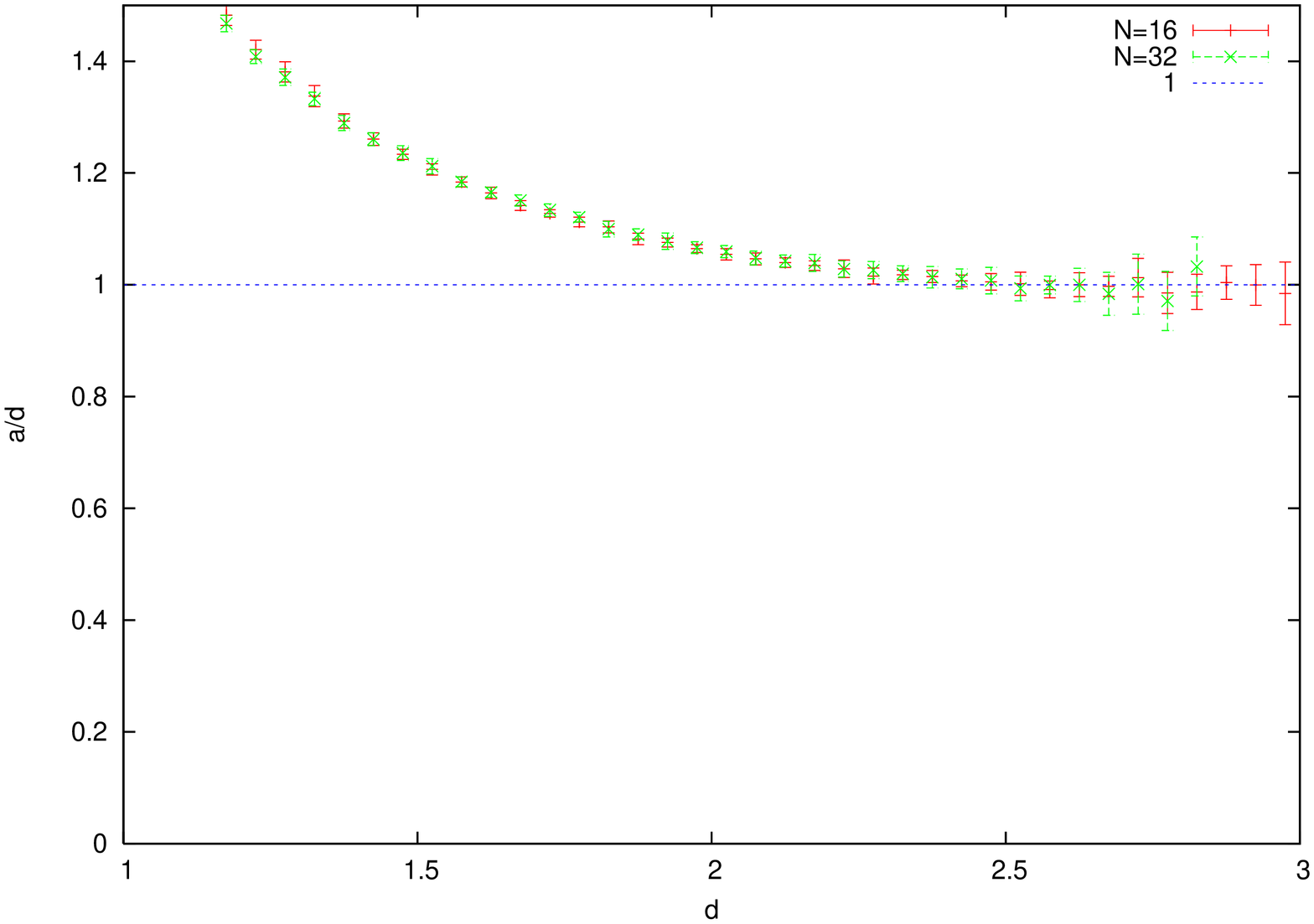}}
  \end{center}
\caption{
$a/d$ vs $d$, $D=4$.  
}\label{fig:Width_D4}
 \end{minipage}
\end{figure}
\begin{figure}[htbp]
 \begin{minipage}{0.45\hsize}
  \begin{center}
  \rotatebox{-90}{
   \includegraphics[width=50mm]{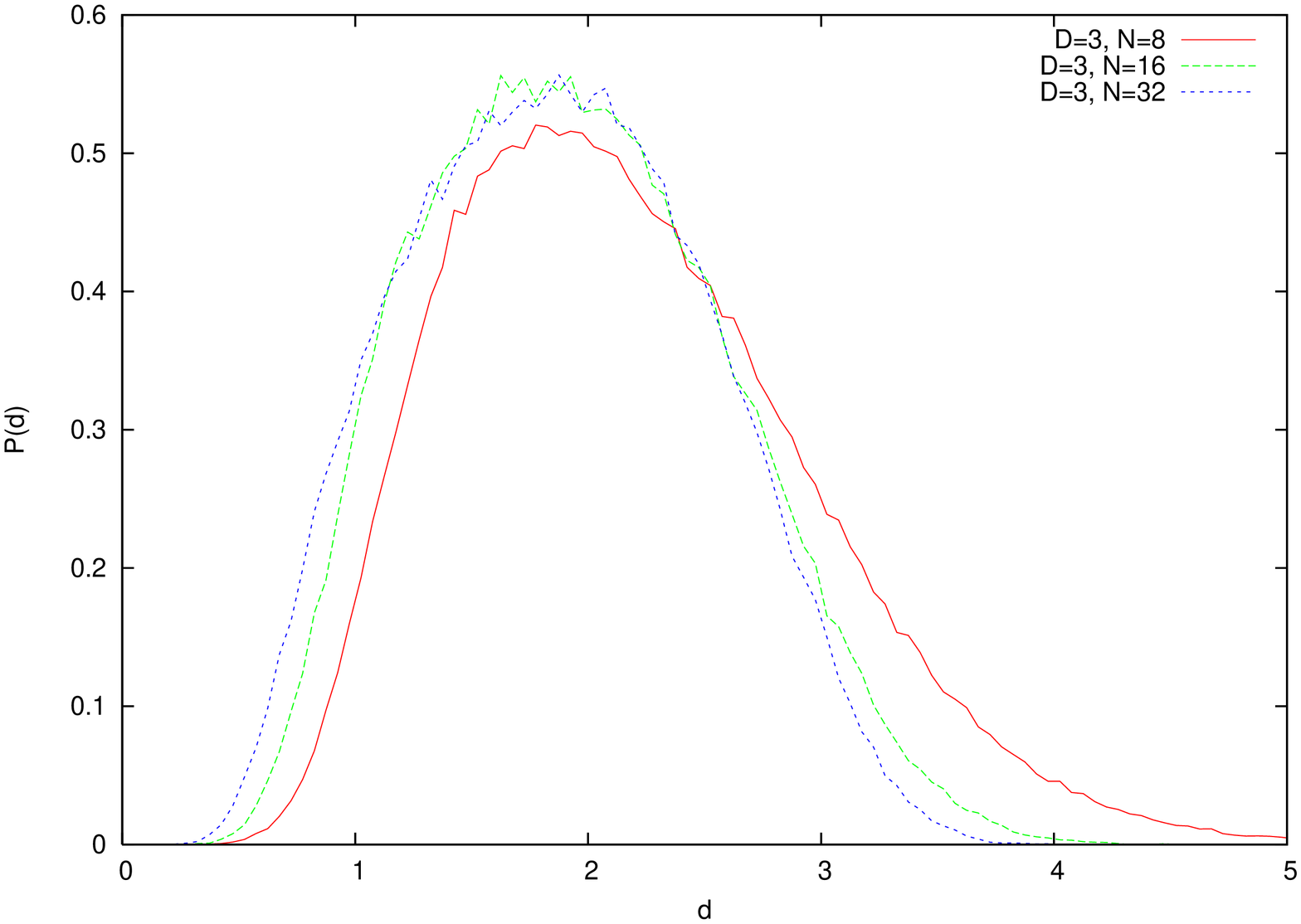}}
  \end{center}
\caption{
Distribution of $d$, $D=3$. 
}\label{fig:DistanceDist_D3}
 \end{minipage}
   \begin{minipage}{0.02\hsize}
  \hspace{-5mm}
  \end{minipage}
 \begin{minipage}{0.45\hsize}
  \begin{center}
  \rotatebox{-90}{
   \includegraphics[width=50mm]{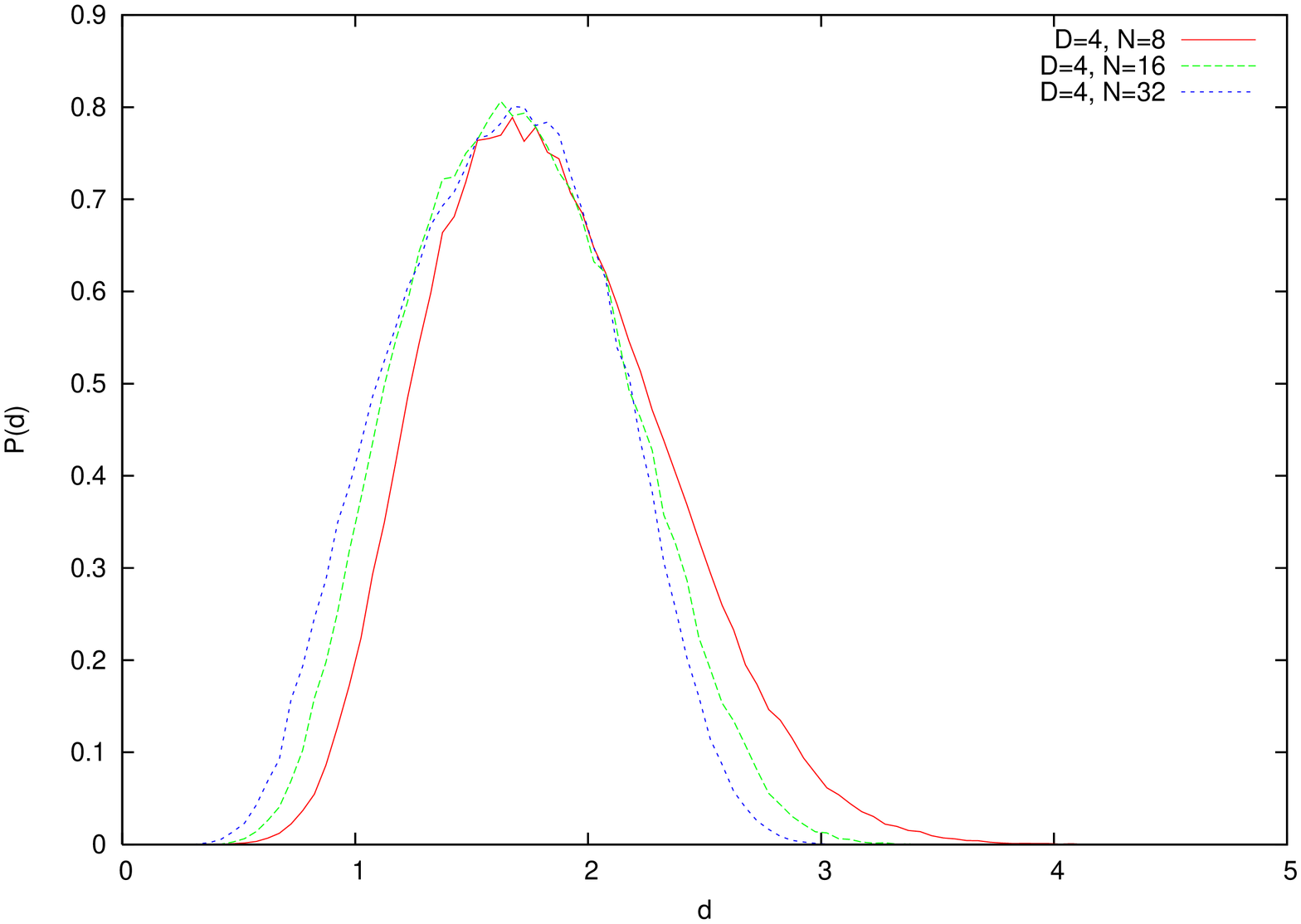}}
  \end{center}
\caption{
Distribution of $d$, $D=4$.  
}\label{fig:DistanceDist_D4}
 \end{minipage}
\end{figure}
As we can see from Fig.~\ref{fig:Width_D3} and Fig.~\ref{fig:Width_D4}, 
(\ref{mass=distance relation}) approximately holds when  
$|\vec{X}_{ii}-\vec{X}_{jj}|\gtrsim 1.5$, which is comparable to
the radius of the D-brane distribution.  
This fact strongly suggests that, at this scale, 
the relative positions of the D-branes can be determined precisely.
In Fig.~\ref{fig:DistanceDist_D3} and Fig.~\ref{fig:DistanceDist_D4} 
we plot the distribution of distances between two points,~\footnote{
For each configuration we have measured distances for $N(N-1)/2$ pairs of points. 
The distances between two points belonging to different configurations are not measured. 
} $P(d)$, 
normalized by the condition $\int dxP(x)=1$, which is consistent with the 
result of \cite{AABHN00}.
We can see that many pairs of branes are 
sufficiently separated so that their relative positions can be determined 
precisely.  
Thus, even in the bosonic model, 
resolution for the D-brane position is rather good 
and the D-brane distribution shown 
in \S~\ref{sec:diag distribution} makes sense. 
This gives enough motivation to apply 
the maximal diagonalization procedure 
to study the BH/BS transition, which will be done in  
\S~\ref{sec:BH/BS transition}. 

As we can see by comparing Fig.~\ref{fig:FitD3N32mass150-155} and 
Fig.~\ref{fig:FitD3N32mass100-105}, 
finite-$N$ corrections are larger at the shorter distance scale. 
This is probably due to the finite-$N$ correction 
coming from the Faddeev-Popov determinant for the maximally diagonal gauge.
We have not evaluated the Faddeev-Popov determinant in full generality. 
For the simple and important case where only one off-diagonal mode is excited, 
the Faddeev-Popov determinant can be evaluated and 
is proportional to 
$\frac{d^4}{4}-\frac{d^2 z^2}{N}+\frac{z^4}{3N^2}$, where 
$d=|X_{ii}-X_{jj}|$, 
$z=\sqrt{|X_{tr}^1|^2+|X_{tr}^2|^2}$ and 
$X_{tr}$ stands for 
the relevant off-diagonal transverse mode of $X^{new}$. 
For large $N$, the first term dominates and one 
can simply neglect the determinant as it only gives a constant factor
when considering the distribution of $z$ for 
fixed $d$.
For finite $N$ and small $d$, $z$ spreads to large values and hence 
the second and the third terms of the determinant should give 
 large contributions.

\section{Matrix quantum mechanics and black hole/black string topology change }
\label{sec:MQM}
\hspace{0.51cm}
The action of the bosonic matrix model in $(0+1)$ dimension is 
\begin{eqnarray}
S
=
N\int_0^L dx\ 
Tr\left(
\frac{1}{2}(DX_i)^2
-
\frac{1}{4}[X_i,X_j]^2
\right). 
\label{action:bosonicMQM}  
\end{eqnarray} 
Here $D=\partial_x-i[A,\ \cdot\ ]$ is the gauge covariant derivative 
and $X_i(i=1,\cdots, d_s)$ are adjoint scalars. 
When $d_s=9$, this model is the high-temperature limit of 
$(1+1)$-dimensional maximal super Yang-Mills theory. 
(We have absorbed 
factors such as $\lambda T_H$ 
in (\ref{ActionBosonicMQMWithoutAbsorption}) 
by redefining fields and the coordinate $x$.)

As is discussed in \S~\ref{sec: gravity dual},
this model is related to the black hole/black string transition. 
We begin with the study of its 
phase structure in \S~\ref{sec:MQM phase structure}. 
Next we generalize the maximal diagonalization procedure to 
higher dimensions in \S~\ref{sec:diagonalization procedure for MQM}. 
Then in \S~\ref{sec:BH/BS transition} we apply this procedure 
to the matrix quantum mechanics and see the topology change. 
We find that, at the transition \cite{AMMW04} corresponding to 
the BH/BS transition \cite{GL93}, 
there is a topology change which strikingly resembles 
the counterpart on the gravity side \cite{KW04}.  
\subsection{Phase structure of matrix quantum mechanics}
\label{sec:MQM phase structure}
\hspace{0.51cm}
In this section we study the phase structure of the bosonic matrix quantum mechanics. 
The argument is parallel to that in \cite{KNT07}; the only difference is 
in the number of scalars, $d_s$. 

As is well-known, the model has a global $U(1)$ symmetry, multiplying 
the Wilson loop 
$TrW\equiv\frac{1}{N}Tr{\rm P}\exp(i\int_0^L dx A_x)$, 
winding on the compactified direction, by a phase factor.
This symmetry can be broken in the large-$N$ limit. In order to detect the breakdown of 
this symmetry, we should measure $\langle|TrW|\rangle$ rather than $\langle TrW\rangle$, 
because the overall phase factor is not fixed at finite-$N$. 
As shown in \cite{AMMW04,KNT07} the global $U(1)$ symmetry is broken 
when $L$ is small. 
To study the detail of the transition, it is better to look at 
eigenvalues of $W$ rather than $TrW$ itself. 
Because $W$ is unitary, its eigenvalues are written as $e^{i\theta}$, 
where $-\pi\le\theta\le\pi$. 
We study the distribution of the phases $\rho(\theta)$. 
Note that we fix the overall phase factor such that $TrW$ becomes real.  
By looking at the distribution $\rho(\theta)$, 
it was found that there are two successive 
transitions at $L=L_1$ and $L=L_2<L_1$ \cite{KNT07}. 
Above $L=L_1$ the phase distribution is uniform, $\rho(\theta)=1/(2\pi)$, 
and hence the $U(1)$ symmetry is not broken, $\langle|TrW|\rangle=0$. 
Therefore, the theory is volume independent \cite{EK82,NN03} in this region. 
Below $L=L_1$ the distribution is not uniform. For $L_2<L<L_1$, $\rho(\theta)$ 
is not a constant and is non-zero everywhere. 
Below $L=L_2$ the distribution has a gap (i.e. $\rho(\theta)$ 
becomes zero at a certain value of $\theta$).  
It has been observed that 
the phase distribution below $L=L_1$ can be fitted \cite{KNT07} 
by an expression used for the Gross-Witten model \cite{GW80}, which is,
\begin{eqnarray}
\frac{1}{2\pi}\left(
1+\frac{2}{\kappa}\cos\theta
\right)
\qquad(\kappa\ge 2)\label{trans1}
\end{eqnarray}
for ungapped phase $L_2\le L< L_1$ and 
\begin{eqnarray}
\frac{2}{\pi\kappa}\cos\frac{\theta}{2}
\sqrt{\frac{\kappa}{2}
-
\sin^2\frac{\theta}{2}}
\qquad(\kappa<2) \label{trans2}
\end{eqnarray}
for gapped phase $L<L_2$. 

\begin{figure}[htbp]  
  \begin{center}
  \rotatebox{-90}{
   \includegraphics[width=40mm]{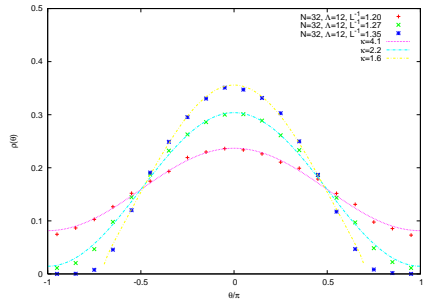}}
  \end{center}
\caption{
Distribution of the Wilson line phase near transition point. $d_s=2, N=32, \Lambda=12$.  
}\label{fig:phase1.eps} 
\end{figure}

In the following we show the results  for $d_s=2$ and $d_s=3$. 
In Figure~\ref{fig:phase1.eps} we show the phase distribution 
for $d_s=2$ when $L$ is very close to $L_2$.
We find the gap emerges at $L_2\simeq (1.3)^{-1}\simeq 0.8$. 
This transition is of second order \cite{KNT07}. 
We note that the corresponding transition in 
the Gross-Witten model is of third order.

In order to determine the value of $L_1$, we utilize 
the $L$-dependence of $\kappa$. For $N=32$ and $\Lambda=12$, 
$\kappa^{-1}$ can be fitted by the straight line, $\kappa^{-1}\simeq 3.3L^{-1}-3.7$, 
at $1.2\le L^{-1}\le 1.26$,   
and we can estimate the value of $L_1$, where $\kappa$ becomes infinity, to be around 
$(1.12)^{-1}\simeq 0.89$.  
To determine the order of the transition, we fit the quantity 
$\int -\frac{1}{N}\langle Tr[X_i,X_j]^2\rangle dx \propto 
\frac{\partial}{\partial L} \log Z$, where $Z$ is the 
partition function, by an ansatz 
\begin{eqnarray}
\int -\frac{1}{N}\langle Tr[X_i,X_j]^2\rangle dx
=
a(L^{-1}-1.12)^p+C, 
\end{eqnarray}
where $C\simeq 0.94$ is the value for $L\ge L_1$.    
Then, 
for the data at $1.2\le L^{-1}\le 1.26$ for $N=32, \Lambda=12$,  
we perform a two parameter fit with respect to $a$ and $p$, and 
we obtain $p\simeq 2.0$, which means that the transition is 
of third order.

For $d_s=3$, the behavior of the eigenvalues are qualitatively the same and, in this case, 
we find $L_1\simeq(0.93)^{-1}\simeq 1.08$ and $L_2\simeq(1.1)^{-1}\simeq 0.9$. 
\subsection{Maximal diagonalization for matrix quantum mechanics: T-dual case}
\label{sec:diagonalization procedure for MQM}
\hspace{0.51cm}
As explained in \S~\ref{sec: gravity dual}, 
we should go to the T-dualized picture in order to see
the 
black hole/black string 
transition.
In the matrix model, we should use 
Taylor's T-duality for Yang-Mills theories~\cite{Taylor96},
which we shall briefly recall here.
The starting point is the zero-dimensional bosonic YM, 
\begin{eqnarray}
S_{0d}=-\frac{1}{4}Tr[X^P,X^Q]^2, 
\end{eqnarray}
where $P, Q=(0,i)=(0;1,\cdots,d_s)$.  
The next step is to  
divide $X^P$ into infinite number of blocks with the size $N$, $N^{1/4}X_{mn}^P$; 
we obtain 
\begin{eqnarray}
S=-\frac{N}{4}tr
\left(
X_{mq}^PX_{qn}^Q-X_{mq}^QX_{qn}^P
\right)
\left(
X_{nr}^PX_{rm}^Q-X_{nr}^QX_{rm}^P
\right), 
\label{Action0Dblock}
\end{eqnarray}
where $tr$ is a trace in each block and the  
summation over $m,n,\cdots$ is assumed. 
Then we impose the compactification constraint, 
\begin{eqnarray}
X_{mn}^i
&=&
X_{(m-1)(n-1)}^i, 
\nonumber\\
X_{mn}^0
&=&
X_{(m-1)(n-1)}^0\qquad (m\neq n), 
\nonumber\\
X_{nn}^0
&=&
X_{(n-1)(n-1)}^0+2\pi R\cdot\textbf{1}. 
\end{eqnarray}
Under this constraint, the action (\ref{Action0Dblock}) should be divided by
the (infinite) number of copies and re-expressed in terms of 
$X_n^P\equiv X_{n0}^P$. For $n\neq 0$ 
these matrices represent open strings between D-branes with the winding number $n$. 
For $n=0$, diagonal components represent positions of D-branes and 
off-diagonal components represent open strings with no winding. 
By T-duality, the winding number is translated to the KK momentum: 
we choose variables after performing the T-dual transformation as  
\begin{eqnarray}
A_x
\equiv
\sum_n e^{2\pi in xR}X_n^0, 
\qquad
Y^i
\equiv
\sum_n e^{2\pi in xR}X_n^i.  
\end{eqnarray}
Then, the action becomes 
\begin{eqnarray}
S_{1d}
=
NR\int_0^{R^{-1}} dx\ tr\left(
\frac{1}{2}(D_xY_i)^2-\frac{1}{4}[Y_i,Y_j]^2
\right). 
\end{eqnarray}
Then, by rescaling 
\begin{eqnarray}
Y^{(new)}=R^{1/3}Y^{(old)},\qquad
A_x^{(new)}=R^{1/3}A_x^{(old)},\qquad
x^{(new)}=R^{-1/3}x^{(old)}
\end{eqnarray}
we obtain 
\begin{eqnarray}
S_{1d}
=
N\int_0^{R^{-4/3}} dx\ tr\left(
\frac{1}{2}(D_xY_i)^2-\frac{1}{4}[Y_i,Y_j]^2
\right). 
\label{action after rescaling}
\end{eqnarray}

From these rules, we learn that  open strings between different D-branes in  
the T-dualized picture correspond to the non-constant modes
and the off-diagonal elements in the one-dimensional YM.
Hence, it is natural to
minimize non-constant modes and off-diagonal elements.
We note that we should distinguish here  
between the minimization
of off-diagonal components (and non-constant modes)
and the maximization
of diagonal components, as one cannot use the 
conservation of the matrix norm by unitary transformations (\ref{RFMatrixNorm}).
(We use the term ``maximal diagonalization" throughout 
this paper although it is a slight abuse of 
the terminology. )  
Actually, the minimization procedure gives sensible results,  
as we will see,
whereas the maximization procedure turns out to give pathological results.
One way to understand the difference is to note
that the measure for the size of off-diagonal (and non-constant) 
elements and diagonal elements are positive. 
For the minimization procedure, this serves as a lower bound to stabilize the
procedure, whereas 
the maximization procedure is more susceptible to pathology 
because of the absence of an upper bound.

Another technical issue is that,
to look at the distribution of branes, we have to 
take care of the effect of the translational symmetry,
and fix the origin for branes. 
To fix the origin for a noncompact direction, we simply take $X$ to be 
traceless.  To fix the origin of the $S^1$ direction, we 
fix the global $U(1)$ symmetry so that the Wilson loop 
$Tr {\rm P} \exp(i\int dx A_x(x))$ becomes real and positive \cite{KNT07}.  

Our simulation is based on the non-lattice technique \cite{HNT07}.  
We generate configurations consisting of
the gauge field in the static-diagonal gauge  
\begin{eqnarray}
A_x=\frac{1}{L}\cdot diag(\alpha_1,\cdots,\alpha_N), 
\qquad
L=R^{-4/3}, 
\end{eqnarray}
and Fourier modes of adjoint scalars 
\begin{eqnarray}
\tilde{Y}_i(p)\ (p=-\Lambda,\cdots,\Lambda). 
\end{eqnarray}
Because the coordinate basis is more convenient for maximal diagonalization,  
we introduce $2\Lambda+1$ lattice sites
\begin{eqnarray}
x
=
-\frac{\Lambda}{2\Lambda+1}L,\ 
\cdots,\ 
0,\ 
\cdots,\ 
\frac{\Lambda}{2\Lambda+1}L,  
\end{eqnarray}
and perform the Fourier transformation  
\begin{eqnarray}
Y_i(x)
=
\sum_{p=-\Lambda}^\Lambda e^{2\pi ipx/L}
\tilde{Y}_i(p).  
\end{eqnarray}
The gauge field is  mapped to the unitary link variable as 
\begin{eqnarray}
U(x)=diag(e^{i\alpha_1/(2\Lambda+1)},\cdots,e^{i\alpha_N/(2\Lambda+1)}) 
\end{eqnarray}
After performing the unitary transformation
in the coordinate basis to maximally diagonalize, 
we have to go back to the momentum basis to use the
T-dualized picture.

From the above considerations, it follows that
one should minimize 
\begin{eqnarray}
\frac{2\Lambda+1}{L^2}\sum_x Tr\left(
2-
U(x)-U^\dagger(x)
\right)
+
\sum_i
\left[
\frac{1}{L^2}
{\rm Re}
\left\{
\sum_x\left(U_{ii}(x)-1\right)
\right\}^2
-
\frac{1}{(2\Lambda+1)^2}
\left(\sum_x Y_{ii}(x)\right)^2
\right]. 
\label{what we should minimize}
\end{eqnarray}
The detailed derivation is given in Appendix~\ref{appendix:diagonalization}.

\subsection{Black hole/black string topology change}
\label{sec:BH/BS transition}
\begin{figure}[htbp]
 \begin{minipage}{0.3\hsize}
  \begin{center}
  \rotatebox{-90}{
   \includegraphics[width=45mm, trim = 0mm 0mm 10mm 0mm,clip=true]{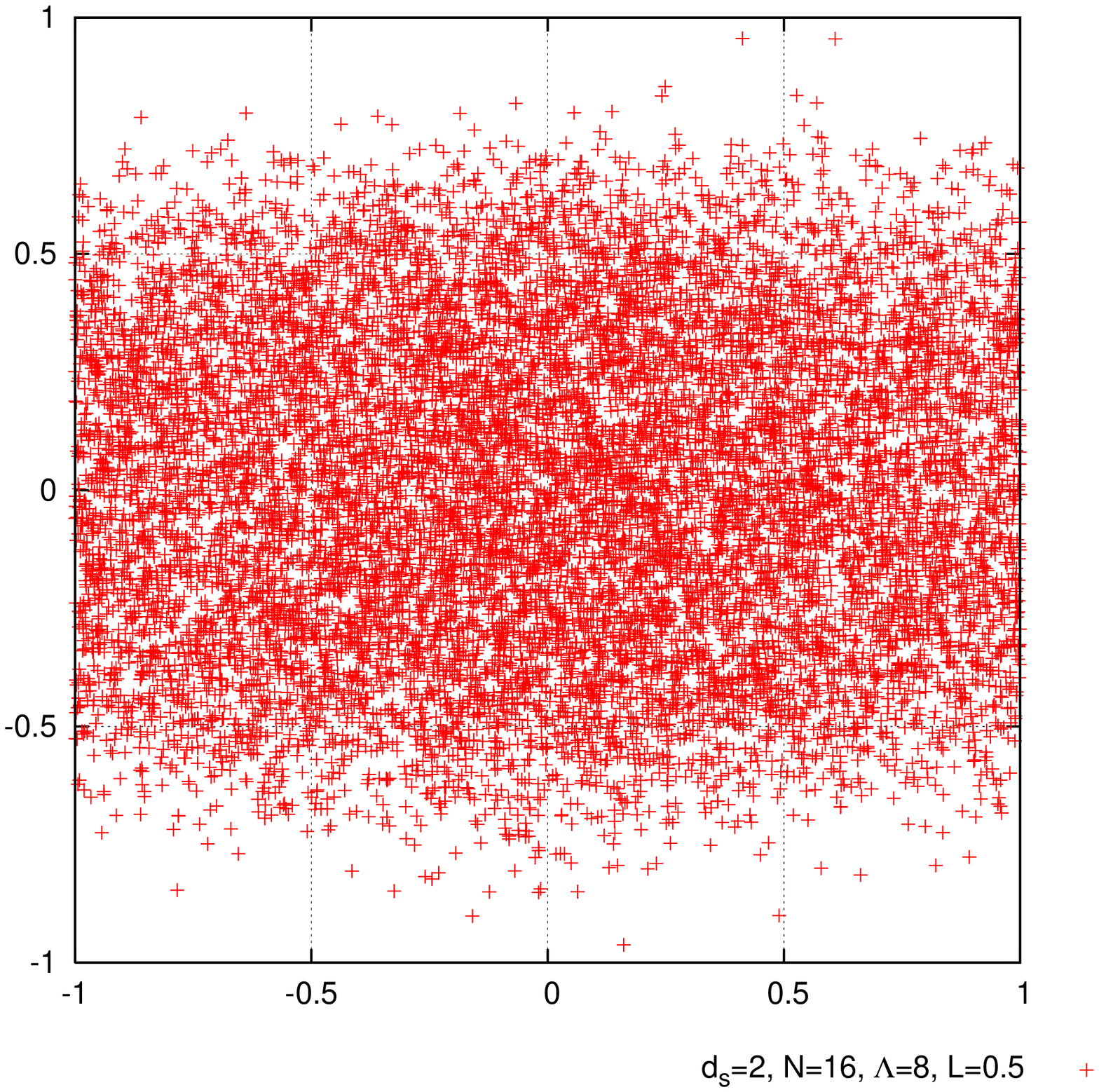}}
  \end{center}
\caption{
Distribution of diagonal components for $d_s=2, N=16, \Lambda=8, L^{-1}=0.5$. 
The ``uniform black string" phase. 
}\label{fig:DiagDistD3N16C8T05}
 \end{minipage}
    \begin{minipage}{0.01\hsize}
  \hspace{-5mm}
  \end{minipage}
   \begin{minipage}{0.3\hsize}
    \begin{center}
  \rotatebox{-90}{
\includegraphics[width=45mm, trim = 0mm 0mm 10mm 0mm,clip=true]{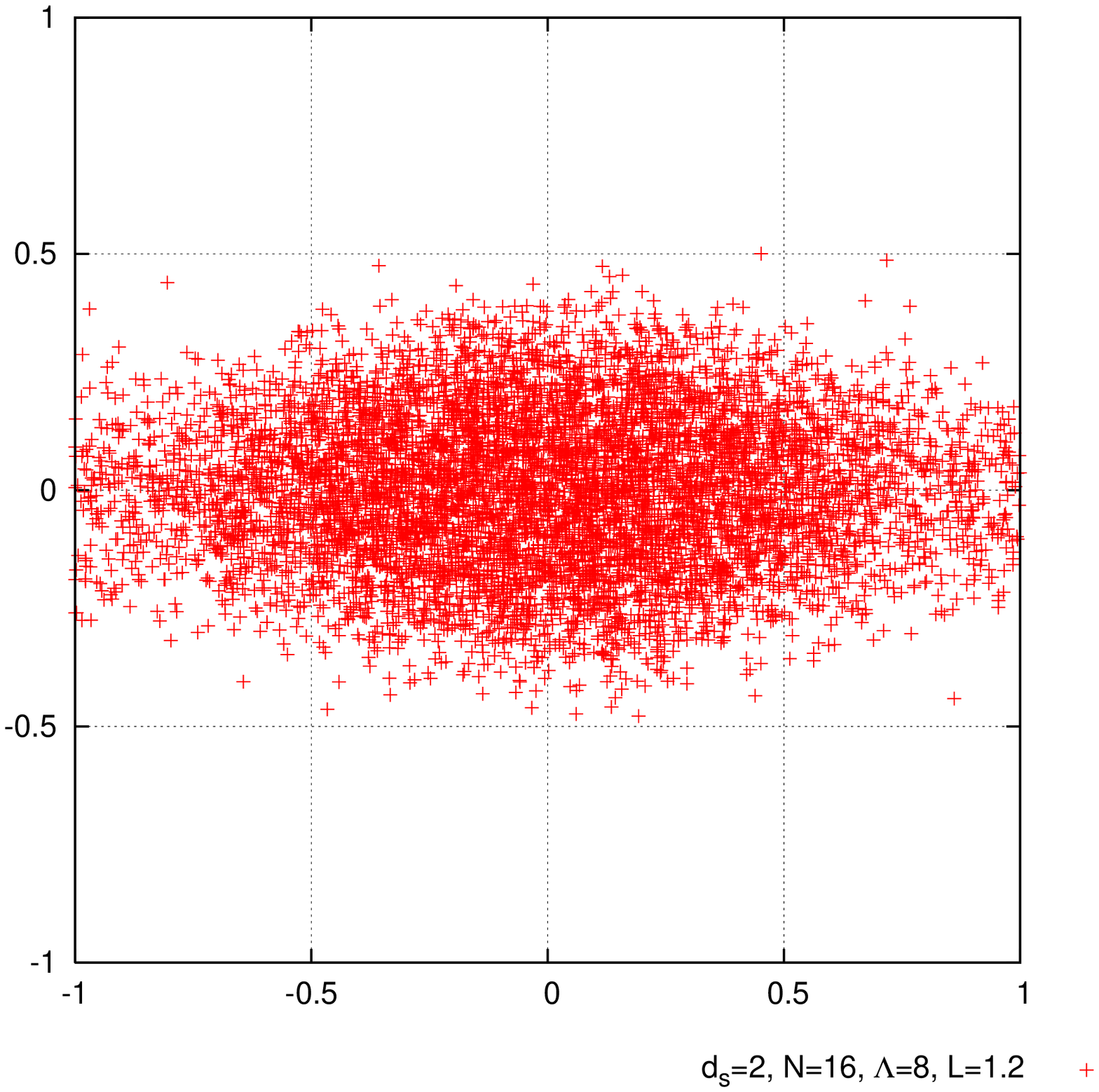}
}
\caption{
Distribution of diagonal components for $d_s=2, N=16, \Lambda=8, L^{-1}=1.2$. 
The ``nonuniform black string" phase.  
}\label{fig:DiagDistD3N16C8T12}
  \end{center}
  \end{minipage}
     \begin{minipage}{0.01\hsize}
  \hspace{-5mm}
  \end{minipage}
 \begin{minipage}{0.3\hsize}
  \begin{center}
  \rotatebox{-90}{
   \includegraphics[width=45mm, trim = 0mm 0mm 10mm 0mm,clip=true]{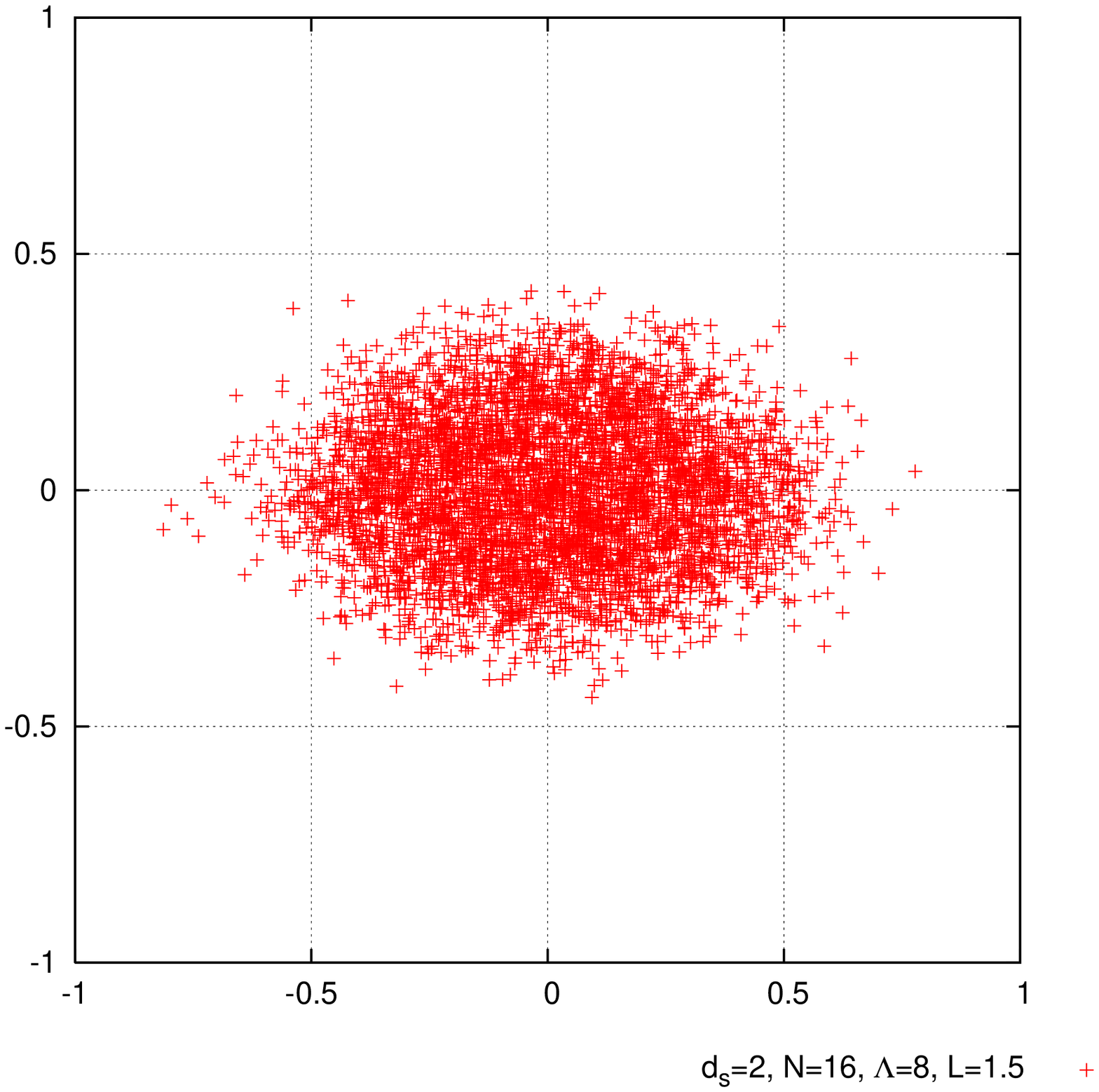}} 
  \end{center}
\caption{
Distribution of diagonal components for $d_s=2, N=16, \Lambda=8, L^{-1}=1.5$. 
The ``black hole" phase.  
}\label{fig:DiagDistD3N16C8T15}
 \end{minipage}
\end{figure}

\hspace{0.51cm}
In this section we show the D-brane distribution, obtained by using 
the maximal diagonalization procedure introduced in the previous subsection.  
The off-diagonal components behaves similarly to
the zero-dimensional model studied in \S~\ref{sec:offdiag distribution} 
and hence the distribution is meaningful.  
In Figure~\ref{fig:DiagDistD3N16C8T05}, \ref{fig:DiagDistD3N16C8T12} and 
\ref{fig:DiagDistD3N16C8T15} 
we plot diagonal components after the maximal diagonalization 
for the $d_s=2$ model. 
The x-axis represents the compactified direction and 
two edges $\pm 1$ should be identified. (In these plots we have rescaled 
the coordinates so that the period of the compactified direction becomes $2$.) 
The y-axis represents the $X_1$ direction. The $X_2$ direction is projected out. 

When the radius $R$ in the T-dual picture is small (i.e. $L$ large) 
they form a uniform string, as we can see from     
Figure~\ref{fig:DiagDistD3N16C8T05}. 
If $R$ becomes large, at some point the string begins to pinch; 
see Figure~\ref{fig:DiagDistD3N16C8T12}.  
We may call it as the nonuniform string. 
Finally it is pinched off completely and becomes a squashed ball, as shown in  
Figure~\ref{fig:DiagDistD3N16C8T15}. 
These shapes are strikingly similar to the black string and the black hole 
in gravity side \cite{KW04}. 
As we increase $R$ further, we have found that 
the distribution exhibits $SO(d_s+1)$ symmetry, 
as expected.  If one does not apply the maximal diagonalization and 
just plots diagonal components resulting from a Monte-Carlo simulation
(in the static-diagonal gauge),
these gradual change of the geometry is almost completely obscured. In particular,
one cannot see the recovery of the spherical symmetry for large $R$.

We have also studied the $d_s=3$ model, and found that the shapes of the bound states 
look similar. The difference is in their internal structures: for $d_s=2$ the bound state is 
uniformly filled ($D^2\times S^1$ for the black string phase and $D^3$ for the black hole phase), 
and for $d_s=3$ the distribution is shell-like 
($S^2\times S^1$ for the black string phase and $S^3$ for the black hole phase).

\section{Supersymmetric matrix models}
\label{sec:SUSY}
\hspace{0.51cm}
In this section we show preliminary simulation results on  
a supersymmetric matrix quantum mechanics.
We adopt the non-lattice Euclidean path-integral method 
\cite{HNT07}. A Hamiltonian formulation can be found in \cite{Wosiek02}, 
and a mean field approximation is studied in \cite{KLL00}. 
As explained in \S~\ref{sec: gravity dual}, at low temperature 
the maximally symmetric model 
(\ref{1dSYMaction}) describes the black 0-brane in IIA supergravity. 
Here we study a simpler model,   
the four-supercharge analogue of (\ref{1dSYMaction}) :  
\begin{eqnarray}
S
=
N
\int_0^{1/T}dt\  
Tr\left(
\frac{1}{2}(D_t X_I)^2
-
\frac{1}{4}
[X_I,X_J]^2
+
\bar{\psi}D_t\psi
-
\bar{\psi}\sigma^I[X_I,\psi]
\right),   
\label{4SUSY MQM action}
\end{eqnarray}
which is obtained from four-dimensional ${\cal N}=1$ SYM through the dimensional reduction. 
Here $I,J$ runs through $1,2$ and $3$, $\sigma^I$ are Pauli matrices, 
and $\psi_\alpha\ (\alpha=1,2)$ are $N\times N$ complex fermionic matrices. 
This model does not suffer from the sign problem and 
the simulation cost is less expensive, 
while we expect its behavior should 
at least qualitatively similar to that of (\ref{1dSYMaction}). 
To consider the finite temperature system we have to impose the anti-periodic boundary condition 
(a.p.b.c.) to the fermion. However, as pointed out in \cite{AHNT07},  
with the a.p.b.c. the system with small $N$
and finite $T$ has an instability (eigenvalues of $X_I$ run to infinity),
which makes the computation heavy as we need to take $N$
to be reasonably large.
This lead us to adopt the periodic boundary condition (p.b.c.), 
as the p.b.c. and a.p.b.c should be the same 
in the $T\to 0$ limit which we are interested in.  

Because the compactified direction is now temporal and the T-duality 
transformation is not performed, 
the maximal diagonalization procedure introduced in 
\S~\ref{sec:diagonalization procedure for MQM} 
is not adequate. Rather, 
we maximize $\sum_I\sum_{p}\int dt \left(X^I_{pp}\right)^2$, 
that is, we maximally diagonalize scalars at each time slice.~\footnote{
A similar maximal diagonalization (and the criterion for its validity),
which does not count the gauge field (or the scalar field)
in the temporal direction exists for 
for the zero-dimensional matrix model
studied in \S~\ref{sec:0dMM}. 
This is more suitable than that given there 
(which treat the gauge field in the temporal direction 
in an equal footing as other scalar fields)
if we regard this system as the high temperature limit 
of the D0-brane system.
The resulting D-brane distributions from these different prescriptions 
do not differ qualitatively. 
}
Numerically, at $N=6$, the diagonal components we find are distributed like a ball. 
To determine the shape of the distribution reliably, it is also necessary to study 
larger $N$ and see the $N$-dependence. We would like to report on this issue in 
a future work. 

We should also introduce a slightly different criterion 
for determining the validity of the maximal diagonalization.
We again regard the diagonal and off-diagonal elements 
to be the position of the D0-branes and open strings, respectively. 
Our criterion is again to see when we can neglect higher order interaction terms
(and accordingly can think of off-diagonal elements as dynamically insignificant).
If interaction terms are negligible, 
and the diagonal elements are slowly-varying compared to 
the off-diagonal elements, each off-diagonal mode 
behaves as a harmonic oscillator.
Quanta of this harmonic oscillator correspond to open strings and
have energies proportional to their lengths. 
As we are interested in the low temperature limit, 
we furthermore require that the open string is not excited at all,
the wave function being of the Gaussian form. 
This is the minimal non-commutativity;
off diagonal components cannot vanish because of the zero point oscillation.

\begin{figure}[htbp]
 \begin{minipage}{0.45\hsize}
   \begin{center}
  \rotatebox{-90}{
   \includegraphics[width=50mm]{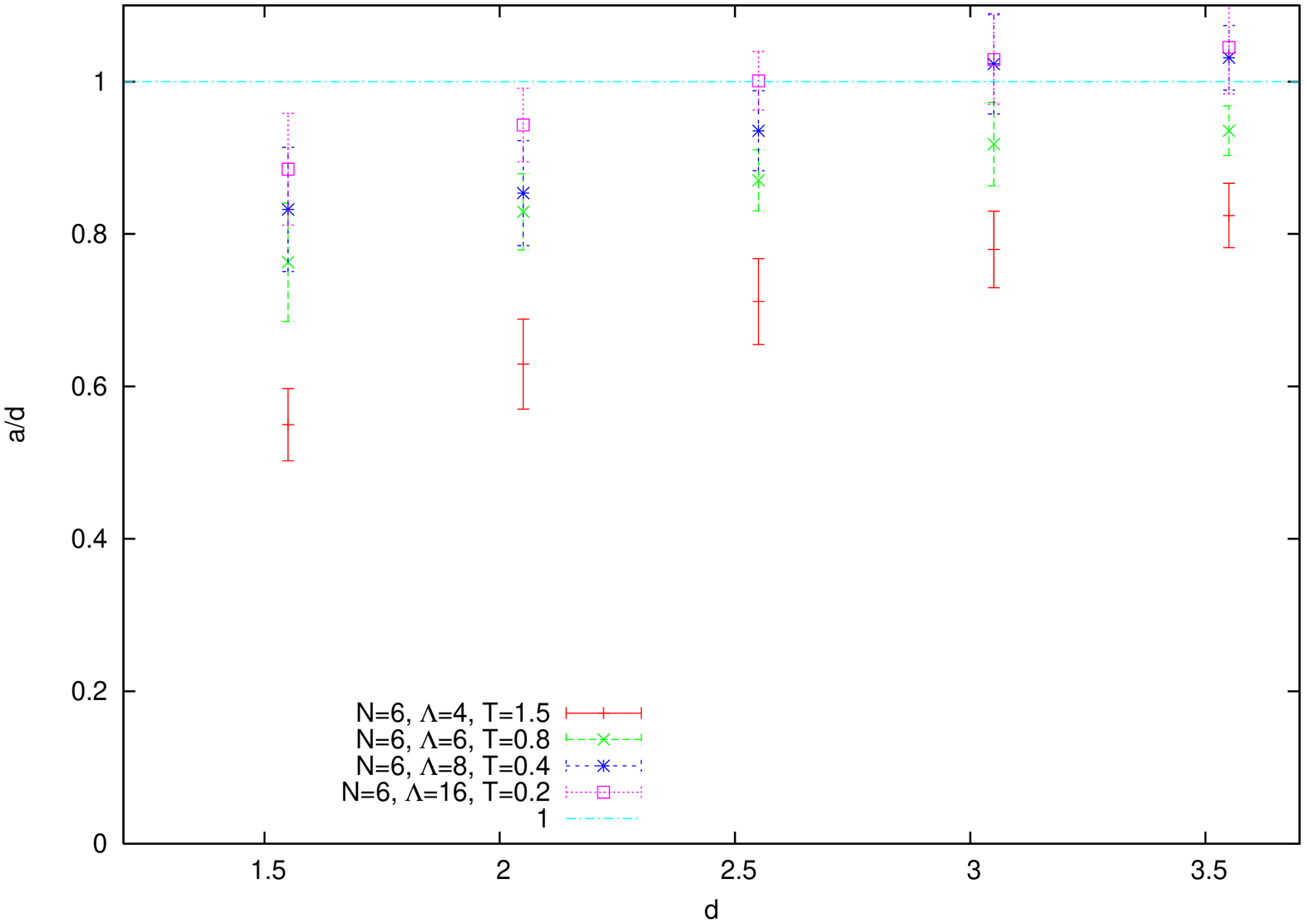}}
  \end{center}
  \caption{
The ratio $a/d$ in 4 SUSY matrix quantum mechanics is plotted against $d$.    
}
  \label{fig:mass vs diatance SUSY MQM}
 \end{minipage}
   \begin{minipage}{0.02\hsize}
  \hspace{-5mm}
  \end{minipage}
 \begin{minipage}{0.45\hsize}
  \begin{center}
  \rotatebox{-90}{
   \includegraphics[width=50mm]{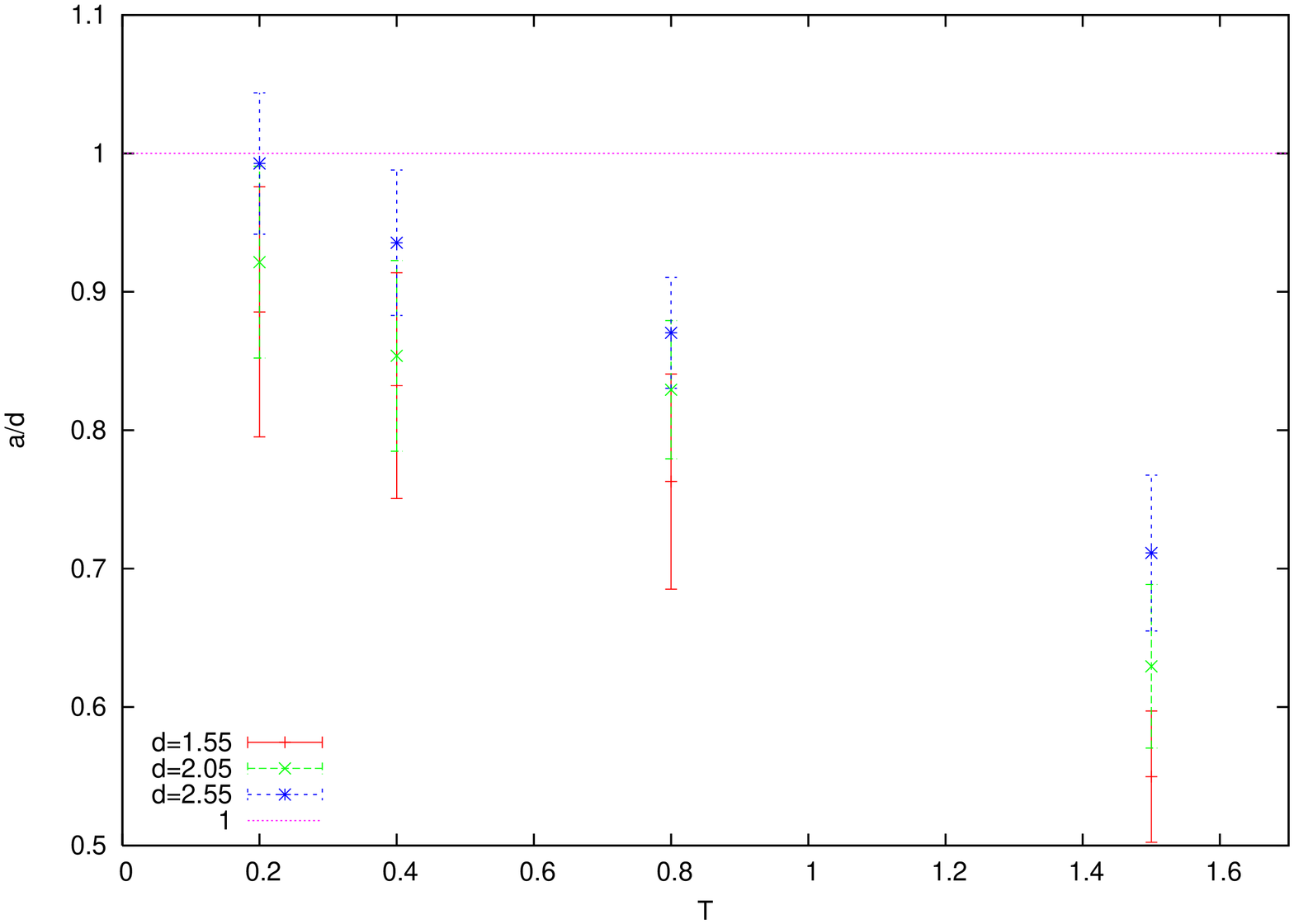}}
  \end{center}
  \caption{ 
  Plot of $a/d$ versus $T$ at $d=1.55, 2.05$ and $2.55$ in 4 SUSY matrix quantum mechanics.  
}
  \label{fig:mass vs diatance SUSY MQM v2}
  \end{minipage}
\end{figure}

Indeed, we find that the distribution of 
$y\equiv\sqrt{N}|X_{ij}|$ can be fitted by 
the probability distribution function
for the zero-point oscillation 
\begin{eqnarray}
\rho(y)=4aye^{-2ay^2}   
\label{harmonic oscillator distribution}
\end{eqnarray}
in a wide region, and $a$ is very 
close to $d\equiv |\vec{X}_{ii}(t)-\vec{X}_{jj}(t)|$. 
In Fig.~\ref{fig:mass vs diatance SUSY MQM} we plot $a/d$ against $d$. 
In Fig.~\ref{fig:mass vs diatance SUSY MQM v2}, we plotted $a/d$ for fixed $d$ 
against $T$. 
At short distance, 
the data begins to depart from the ansatz (\ref{harmonic oscillator distribution}).
We suspect the reason to be 
corrections in the Faddeev-Popov determinant,
which are not negligible for finite $N$
as we have seen in \S~\ref{sec:offdiag distribution};  
to study $d\lesssim 1.5$ reliably, we have to take larger $N$. 

The applicability of the ansatz (\ref{harmonic oscillator distribution}) 
is at least an indication of the validity of 
the assumption that 
the D0-branes (diagonal components) are more slowly-varying than 
the open strings (off-diagonal components). 
We note that this assumption is crucial for previous studies~\cite{DFS96,KS99} 
for D-brane bound states, using various analytic and approximate methods.
For another approach to bound states, see \cite{Witten82,BS86,HP00,Yi:1997eg}. 
For a discussion of bound states for the zero-dimensional matrix model, 
see \cite{AIKKT98}.

\begin{figure}[tbp]
\begin{center}
\scalebox{0.25}{
\rotatebox{-90}{
\includegraphics[width=20cm,height=30cm]{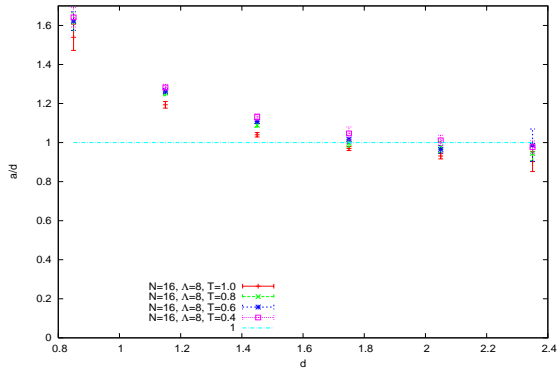}
}}
  \caption{ 
 The ratio $a/d$ in the bosonic matrix quantum mechanics 
 with 3 scalars is plotted against $d$. }
  \label{fig:mass vs diatance bos MQM} 
\end{center}
\end{figure}

From these plots, 
we can see that the notion of the D-brane distribution becomes better 
at smaller $T$.~\footnote{ 
This behavior is in sharp contrast with that in the bosonic model shown in 
Fig.~\ref{fig:mass vs diatance bos MQM}. 
In the bosonic case, the temperature 
dependence disappears at low temperature.
This is as expected, because for the bosonic model,
the global $U(1)$ symmetry (multiplying the Wilson loop by a phase factor) 
is recovered at low temperature,
which implies that the large-$N$ dynamics does not
depend on the volume (in this case the temperature) ~\cite{EK82,NN03}, 
as is well known. 
The consistency with this general result
makes us rather confident that we are taking a sufficiently large $N$
to neglect finite-$N$ effects. 
In the supersymmetric case the global $U(1)$ symmetry 
is recovered only at $T=0$ \cite{HMN08} and hence the dependence of $T$
does not disappear.}
A more interesting possibility 
suggested by these plots is that $a/d=1$ holds for any $d$ at $T=0$
in the planar limit.  
Let us elaborate on this possibility and the general tendency that
our criterion holds better at lower temperature.
As we are taking the $T\rightarrow 0$ limit, it is expected
that the off-diagonal elements are in ground states, 
not in excited states.
What is surprising is that the effect
of the higher-order interaction terms seems to be negligible
even for small distances between D-branes,
because for small distances the width of the wave function
(for each off-diagonal element)
is large and therefore one expects that the higher order terms
in the action also contribute.
The fact that this absence of the effect of interaction terms
occurs for supersymmetric models and not for bosonic models
suggests that this might be a consequence of the familiar cancellation
between bosonic and fermionic degrees of freedom.
It is a very interesting problem to understand the mechanism
of this phenomenon.
Of course, to obtain a rigid conclusion we also have to study the model
more thoroughly for larger $N$ 
and smaller $T$. We would like to report on these issues
in a future publication.  

It might be worthwhile to compare this with an approach 
\cite{Berenstein05,BerensteinEtAl}, 
centering around the Berenstein's conjecture \cite{Berenstein05} 
on the strongly coupled 
${\cal N}=4$ SYM on ${\mathbb R}\times S^3$.
As this approach focuses on
dynamics of the diagonal elements, neglecting 
that of the off-diagonal elements,
it is closely related to the themes of this paper. 
In this approach one has an additional dimensionful parameter, 
the mass for scalar fields, originating from the curvature of the $S^3$.
Because of this, under a crucial assumption explained
below, 
the strong coupling limit implies very large
mass terms for the off-diagonal elements.
This in turn implies that the width of the wave function
(or the Gaussian) is very small, which automatically validates
the neglection of interaction terms.
This is similar to, but has a crucial difference from 
what seems to be happening in our model:
for our models, the widths of the Gaussian distributions are generated 
dynamically, and are finite instead of vanishingly small. Nonetheless, 
also in our model higher order interaction terms
seem not to have effect in dynamics in the low temperature regime.

The assumption mentioned above, underlying the analysis of
${\cal N}=4$ SYM on ${\mathbb R}\times S^3$,
is that
the typical length scale for distances between D-branes
should be determined basically by the additional dimensionful parameter
(the size of $S^3$) and should not depend too much on the coupling constant,
which is also dimensionful. 
In \cite{BHH08}, a mechanism for explaining
this behavior is proposed, which is also closely related
to the behavior we find numerically for $T\rightarrow 0$. 
The mechanism is based on an instability
of the D-brane bound state in finite-$N$ supersymmetric matrix models 
(without mass-terms for scalars), 
where distances between D-branes become arbitrarily large.~\footnote{
This instability, which is of interest by itself, is suppressed severely
for large $N$; the bound state we have discussed in this section 
might be called as metastable in this sense. 
This instability is numerically found in \cite{AHNT07}, and presumably is 
the same as that first discussed in \cite{dWLN89}.
}
In the presence of the mass-term for scalars, there would be
a competition between this instability and the attractive force from
the mass-term. The balance between these competing effects
would be achieved at the point where the coordinates of D-branes 
take values of the order of the length scale determined by the mass-term,
thus validating the assumption. 
One can partially justify the existence of this instability by using the one-loop
approximation~\cite{CW07, HKN06} which is valid once the distances
between D-branes become large. 
Our numerical results suggest that one might be able to extend this
result even for shorter distances.

As also noted above, the low temperature behavior we find numerically, namely
the absence of the effect of higher order interaction terms for 
a rather short distance scale,
suggests that physics in the low temperature regime might be understood
by an one-loop approximation.  
We hope that our numerical results provide a guide to such an 
improved perturbation theory.

\section{Conclusion and Discussions}\label{sec:discussion}
\hspace{0.51cm}
In this paper we have proposed a simple procedure to determine the shape 
of the D-brane bound state from super Yang-Mills theories. 
The strategy is to maximally diagonalize the matrices representing 
the collective coordinates of the D-branes.  
We have introduced a criterion which determines when the notion 
of this D-brane position makes sense. 
We have tested the use of these ideas by Monte-Carlo simulations.
As the maximally supersymmetric SYM are computationally hard, 
we have studied simpler models-- zero-dimensional and one-dimensional bosonic YM and 
one-dimensional SYM with four supercharges.  
In the bosonic models, the criterion is satisfied 
for the typical distance scale between D-branes. 
The size of the whole distribution of D-branes is a few times larger than 
this length scale below which one cannot talk about D-brane positions. 
This is natural because a theory of quantum gravity would have a minimal length scale 
and very hot black holes, presumably described by the bosonic matrix models, 
would be objects of the size comparable to this scale. 
We also observed the topology change of the D-brane distribution 
corresponding to the BH/BS transition, by combining the above procedure 
and the T-duality prescription for YM. 
In the supersymmetric model, the notion of the D-brane position seems to 
be precise even at shorter distance scales. 

There are many future directions. 
Most straightforward one is to try to understand the gauge/gravity correspondence,  
and furthermore, to study the black hole physics via the correspondence. 
We have already seen the analogue of the BH/BS transition in the bosonic model. 
It is in principle 
straightforward to repeat the similar analysis for (1+1)-dimensional SYM. 
The lattice formulation 
for (1+1)-dimensional SYM has been developed recently \cite{2dlatticeSYMformulation} 
and already Monte-Carlo simulations
have been performed for four-supercharge $SU(2)$ gauge theory 
\cite{2dlatticeSYMsimulation}. 
By carrying out simulations for larger $N$ and 
larger supercharges the BH/BS transition can be directly studied. 
At low but finite temperature, we expect 
to see even the stringy correction to the transition. 
As we have reviewed in \S~\ref{sec: gravity dual}, 
the pattern of the phase transition changes at intermediate temperature -- 
although there are two successive phase transitions at high temperature, 
the dual gravity analysis predicts there is only one transition at low temperature. 
It should be possible to see how high and low temperature regions are interpolated. 
At present it is still numerically challenging, but will be possible in near future.  
Also it will be interesting to study (1+2)- and (1+3)-dimensional SYM 
compactified on torus, which are expected to have richer phase diagrams \cite{HN07,NNR07}.  
A generalization to curved spacetime is also interesting.  
For this purpose, the T-duality procedure on curved spacetime \cite{IIST07} and 
a construction \cite{IIST08} of the four-dimensional ${\cal N}=4$ planar SYM on $S^3$ 
from the BMN matrix model \cite{BMN02} might be useful.

Another interesting direction is to apply our techniques 
to study noncommutative spaces  
and supermembranes \cite{dWHN88}.   
As we can see from Figure \ref{fig:fuzzy sphere spin 250},  
the maximal diagonalization procedure can be a powerful tool to read off the shape 
of noncommutative manifolds. 
In this case,  off-diagonal components are 
very small, although non-zero, and the shape is determined precisely.   
For example, we might be able to see the 
shape of the higher-genus membranes (fuzzy surfaces) 
constructed along the line of \cite{ABHHS06}. 
For the case of membranes, the embedding of the topology 
in the matrices has been discussed in \cite{Shimada03}, 
and it will be interesting to see the relation with the current approach. 
The study of the ground state of the matrix models for string/M theory 
\cite{dWHN88,BFSS96,IKKT96}, which presumably is associated with a different type of 
large $N$ limit than the planar limit, is also an important direction. 

We wish to stress that the rather high applicability of
the Gaussian distribution 
for the off-diagonal elements are, although natural, far 
from trivial. 
This Gaussian behavior suggests an existence of an
improved perturbation theory 
(presumably some kind
of mean-field approximations incorporating the
information of the distribution of the diagonal elements)
and its validity at the one-loop level.
Usually one does not expect that the perturbation theory
should be useful, as the 't Hooft coupling constant $\lambda$
is dimensionful and we cannot compete it with 
the kinetic term for the mass-less case we are interested in.
(For matrix models with a mass-term 
the ratio between the mass and the 't Hooft coupling constant
gives a good estimate about the applicability of the perturbation scheme.~\footnote{
One might think that the dimensionful parameter $T$ can be used to
construct a dimension-less expansion parameter.
Although this is useful for some systems, 
the expansion parameter is relevant only for the decoupling of 
Kaluza-Klein modes in the temporal direction,
and does not validate the usual perturbation expansion
based on Gaussian integrals for our matrix models without 
the mass terms. 
In \cite{KNT07_HTE}, a hybrid approach, in which KK modes are integrated out 
perturbatively and zero modes are treated by Monte-Carlo simulations, 
is studied.  
})
It is also interesting that the validity of the perturbation
is associated with each pair of D-branes 
rather than with the whole configuration. 
As shown in \S~\ref{sec:SUSY}
for the supersymmetric matrix quantum mechanics at low temperature,
the one-loop approximation is valid for very short distance scales, 
and the one-loop effective action provides a precise description.  
This observation may support a proposal in \cite{Smilga08}, that 
a critical exponent $2.8$ of the energy density of the black hole, 
$\frac{E}{N^2}\sim 7.4 T^{2.8}$, 
predicted by using the supergravity dual \cite{IMSY98}, 
might be understood from the one-loop effective action \cite{OY98}. 
It will be nice if we can reproduce not only the exponent 
but also the overall factor $7.4$.  
More importantly, the distribution of D-branes 
given by our method should correspond somehow
to the metric of the black hole. Clarification of 
this correspondence would give us a good clue
to understand the gauge/gravity duality directly.
We hope that our method provides a useful mean to organize data and
extract physics from it.

\centerline{\bf Acknowledgments} 
The authors would like to thank O.~Aharony, J.~Hoppe, H.~Kawai, Y.~Kimura, 
T.~McLoughlin, L.~Mannelli, J.~Nishimura, 
T.~Nishioka, N.~Obers, D.~Robles-Llana, S.~Theisen and D.~Yamada for discussions and comments. 
T.~A. and T.~H. would like to thank the Japan Society for the Promotion of Science
for financial support. 
M.~H. would like to thank also for Niels Bohr Institute, Albert Einstein Institute, 
University of Tokyo Hongo and RIKEN Nishina Center  for 
hospitality during his stay. 
H.~S. would like to thank IHES and University of Tokyo Komaba for hospitality. 
H.~S. is supported by the grant SFB647 ``Raum-Zeit-Materie''.  
The numerical computation in this work were  
carried out at the Yukawa Institute Computer Facility.

\appendix
\section{Details of the maximal diagonalization in matrix quantum mechanics 
in T-dual picture
}\label{appendix:diagonalization}
\hspace{0.51cm}
In this appendix we explain why we should minimize (\ref{what we should minimize}) 
in the maximal diagonalization procedure. 

Because 
\begin{eqnarray}
U(x)={\rm P}\exp\left(i\int_x^{x+a} dx' A(x')\right)\simeq e^{iaA(x)},
\qquad
U(x)+U^\dagger(x)\simeq 2-a^2 A(x)^2,  
\end{eqnarray}
we have 
\begin{eqnarray}
\frac{1}{a^2}\sum_x Tr\left(
2-
U(x)-U^\dagger(x)
\right)
\simeq
\sum_x Tr A(x)^2
=
(2\Lambda+1)
\sum_p\sum_{ij}\left|\tilde{A}_{ij}(p)\right|^2. 
\end{eqnarray}
Here $a=L/(2\Lambda+1)$ is the lattice spacing. 
Also, the zero-mode of diagonal components can be approximated as 
\begin{eqnarray}
\tilde{A}_{ii}(0)
=
\frac{1}{L}\sum_x a A_{ii}(x)
=
\frac{-i}{L}\sum_x\left(U_{ii}(x)-1\right). 
\end{eqnarray}
Therefore, the contribution to the quantity we wish to minimize 
from the off-diagonal and nonzero modes can be written as 
\begin{eqnarray}
\sum_p\sum_{ij}\left|\tilde{A}_{ij}(p)\right|^2
-
\tilde{A}_{ii}(0)^2
\simeq
\frac{2\Lambda+1}{L^2}\sum_x Tr\left(
2-
U(x)-U^\dagger(x)
\right)
+
\frac{1}{L^2}\sum_i
{\rm Re}
\left\{
\sum_x\left(U_{ii}(x)-1\right)
\right\}^2. 
\end{eqnarray}
Expressions for adjoint scalars are also similar. 
We have 
\begin{eqnarray}
\sum_x Tr Y(x)^2
=
(2\Lambda+1)
\sum_p\sum_{ij}\left|\tilde{Y}_{ij}(p)\right|^2
\end{eqnarray}
and 
\begin{eqnarray}
\tilde{Y}(0)=\frac{1}{2\Lambda+1}\sum_x Y(x), 
\end{eqnarray}
and hence the total contribution from adjoint scalar is 
\begin{eqnarray}
\sum_p\sum_{ij}\left|\tilde{Y}_{ij}(p)\right|^2
-
\sum_{i}\left|\tilde{Y}_{ii}(0)\right|^2
=
\frac{1}{2\Lambda+1}\sum_x Tr Y(x)^2
-
\sum_i\left(\frac{1}{2\Lambda+1}\sum_x Y_{ii}(x)\right)^2. 
\end{eqnarray}
As $Tr Y(x)^2$ is gauge invariant, we need to 
evaluate only the second term. 
In summary, we should minimize 
\begin{eqnarray}
\frac{2\Lambda+1}{L^2}\sum_x Tr\left(
2-
U(x)-U^\dagger(x)
\right)
+
\sum_i
\left[
\frac{1}{L^2}
{\rm Re}
\left\{
\sum_x\left(U_{ii}(x)-1\right)
\right\}^2
-
\frac{1}{(2\Lambda+1)^2}
\left(\sum_x Y_{ii}(x)\right)^2
\right]. 
\end{eqnarray} 
Note that first term does not appear if we perform 
the maximization of diagonal components instead of 
the minimization of off-diagonal components. 


\end{document}